\documentclass[preprint,prd,amssymb,amsmath,nobibnotes,nofootinbib]{revtex4}
\usepackage{amsfonts}
\usepackage{revsymb}
\usepackage{graphicx,epsfig}
\usepackage{placeins}

\begin{document}
\title {Second-order gravitational self-force}
\author{Eran Rosenthal}
\affiliation{
Department of Physics, University of Guelph, Guelph, Ontario  N1G 2W1,
Canada}

\date{\today}

\begin{abstract}
We derive an expression for the second-order gravitational self-force that
 acts on a self-gravitating compact-object moving in a curved background 
spacetime.
First we develop a new method of derivation and apply it to the derivation 
of the first-order gravitational self-force. 
Here we find that our result 
conforms with the previously derived expression.
Next we generalize our method and derive 
 a new expression for the second-order  gravitational self-force. 
This study also has a practical motivation:
The data analysis for the planned gravitational wave detector LISA requires
 construction of waveforms templates for the expected gravitational waves.
Calculation of the two leading orders of the gravitational self-force
 will enable one to construct highly accurate waveform
templates, which are needed 
for the data analysis of 
gravitational-waves that are emitted from extreme mass-ratio binaries.
\end{abstract}

\maketitle

\section{Introduction}

When a self gravitating compact object moves in curved spacetime 
it interacts with its own gravitational field. This interaction 
induces a (gravitational) self-force that
the object exerts on itself, and thereby affects its own motion. 
The analysis of this phenomenon is much simplified  
by considering a point particle
limit, in which the spatial dimensions of the object
approach zero. In this limit, the motion of the object 
is described by a timelike worldline, and the dynamics of the object is 
governed by an equation of motion.

To derive an expression for the self-force,
it is useful to employ perturbations methods. 
In this approximation, one assumes that the object self-gravity 
can be represented by small metric perturbations on a given curved background metric. 
At the leading order of approximation, the worldline of the 
object traces a geodesic in the background spacetime  (see e.g. \cite{DEATH}).
At higher orders, the metric perturbations produced by the object, give rise 
to a self force that induces a deviation from the geodesic trajectory.
The linear metric perturbations that 
are proportional to the mass of the object $\mu$  
induce a first-order self-force --  proportional to $\mu^2$  \cite{MST,QW}.
At the next order, the second-order metric perturbations that are proportional to $\mu^2$,
give rise to a second-order self-force --  proportional to $\mu^3$.

The main difficulty in deriving an expression for 
the self-force originates 
from the fact that we consider a point particle limit. 
In this limit, the metric perturbations become
singular at the location of the particle, and one is required 
to introduce a regularization method to be able to 
derive  the correct (and finite) expression for self-force.
Using different methods Mino Sasaki and Tanaka \cite{MST}, 
and independently Quinn and Wald \cite{QW} have derived 
the general expression for the first-order gravitational 
self-force in a vacuum background
spacetime. To be able to go beyond the first-order self-force and 
calculate the trajectory of 
an object to a higher accuracy, one is required 
 to calculate the interaction of the object with its own
 second-order metric perturbations. These perturbations 
satisfy perturbations equations with highly 
singular source terms.
Recent studies have introduced a regularization method that enables  
one to solve these equations and construct the second-order metric
perturbations  \cite{eran1,eran2}. In this article we follow Refs. \cite{eran1,eran2} 
and use the second-order metric perturbations to construct an expression 
for the second-order gravitational self-force.

The study of self-force has attracted much attention in recent years (see e.g. \cite{PR} and
references therein) mostly because a calculation of the self-force is 
needed for future detection of gravitational waves with the planned 
laser interferometer space antenna (LISA) \cite{LISA}. 
One of the most interesting sources for LISA 
are extreme mass ratio binaries, in which a compact object (e.g. 
a neutron star or a stellar mass black hole) gradually inspirals towards 
a supermassive black hole with a mass $M$ (e.g., $M/\mu= 10^{5}$).  
Calculating the self-force that the compact object exerts on itself enables one to 
calculate the inspiral trajectory, and thereby prepare gravitational 
waveform templates for the expected gravitational waves.
These templates are needed for matched-filtering data-analysis techniques. 
To determine the binary parameters using  
matched-filtering techniques one is often required  
to prepare  gravitational waveform templates with a phase 
error of less than one-cycle over a year of inspiral \cite{BC}.
Calculating waveform templates to this accuracy 
is a challenging task and requires one to take into account the
interaction of the compact object with its own
second-order metric perturbations \cite{eran1,eran2}. This provides a practical motivation 
for the study of the second-order gravitational self-force.

Our analysis is based on perturbation theory.
We therefore decompose the full spacetime metric $g^{full}_{\mu\nu}$ as follows 
\begin{equation}\label{fullg}
g^{full}_{\mu\nu}=g_{\mu\nu}+\delta g_{\mu\nu}\,.
\end{equation}
Here $g_{\mu\nu}$ denotes the 
background metric satisfying the vacuum Einstein's field
equations in the absence of the compact object, and 
$\delta g_{\mu\nu}$ denotes the metric perturbations 
produced by the compact object. Throughout we shall use $g_{\mu\nu}$ to raise and 
lower tensorial indices and to evaluate covariant derivatives.
Recall that the gravitational self-force originates 
from the interaction of the object with its own metric perturbations.
Therefore, the dynamics of the object 
should be described by an equation of motion of the following form
\begin{equation}\label{eom}
\mu \frac{D^2 z^\mu}{D\tau^2}=f^\mu[g_{\mu\nu},\delta g_{\mu\nu},z(\tau)]\,,
\end{equation}
where  $f^\mu$ denotes the desired gravitational 
self-force \footnote{Here it is understood that the dependence on $z(\tau)$ is other 
then terms of the form $C\frac{D^2 z^\mu}{D\tau^2}$ where $C$ is a constant.
No observation can distinguish between these terms 
and the term on the left hand side. Therefore, these terms should be absorbed within
 the term on left hand side in a mass redefinition procedure.},
 $z(\tau)$ denotes the worldline of the object in the background geometry (at the 
point particle limit), and
$\tau$ denotes the proper time with respect to the background geometry. 
Recall that the perturbations  $\delta g_{\mu\nu}$ are singular 
at the worldline. This singularity suggests that 
$f^\mu[g_{\mu\nu},\delta g_{\mu\nu},z(\tau)]$ 
should be interpreted as a regularization operator, which takes 
singular quantities in its input and produces a regular self-force as an output.  
To make this statement meaningful (and useful), we assume that $f^\mu$ depends on the values 
of  $\delta g_{\mu\nu}$ in the vicinity of the worldline.
More precisely, we consider the hypersurface 
of constant time $\Sigma(\tau)$ generated by spacelike geodesics  
 normal to the worldline at $z(\tau)$, and assume that the dependence of 
$f^\mu(\tau)$ on the metric perturbations is in fact a dependence 
on $\delta g_{\mu\nu}[\Sigma(\tau)]$ in the vicinity of $z(\tau)$.
Throughout this article  we shall  ignore 
  corrections to the self-force which originate 
from finite size and asphericity of the object. 
These corrections could be analyzed separately using other methods (see
e.g. \cite{dixon,harte}). For sufficiently spherical objects 
these terms do not affect the first two 
leading orders which concern us here 
\footnote {In a simpler case of 
an electromagnetic self-force in flat 
spacetime it  was recently demonstrated that the self-force is 
universal - i.e. at the point 
particle limit it is independent of the object 
charge distribution \cite{OR1,OR2}.}.

We formally expand the metric perturbations in powers of $\mu$ which gives
\begin{equation}\label{deltag}
\delta g_{\mu \nu}=\mu h_{\mu\nu}+\mu^2 l_{\mu\nu}+O(\mu^3)\,.
\end{equation}
Throughout this paper we keep the dependence on $\mu$ explicit. 
In the context of perturbations theory we can separate
 the calculation
of the self-force into several consecutive steps. In the first step, we completely
 ignore the self-force and consider a geodesic worldline $z_G(\tau)$.
Using  $z_G(\tau)$ as a leading order worldline, 
we can calculate the first-order metric perturbations 
$h_{\mu\nu}$. In this context the first-order self-force is calculated 
using $h_{\mu\nu}$ and $z_G(\tau)$. The corrections to $z_G(\tau)$ due
to the first-order self-force are required as an input for the higher orders terms 
of the self-force,   
for example they are required for the calculation of  the second order self-force.
Formally expanding $f^\mu$ in powers of $\mu$ we obtain 
\begin{equation}\label{foso}
f^\mu=
\mu^2 f^\mu_{(1)}[g_{\mu\nu},h_{\mu\nu},z_G(\tau)]+
\mu^3 f^\mu_{(2)}[g_{\mu\nu},h_{\mu\nu},l_{\mu\nu},z(\tau)]+O(\mu^4)\,.
\end{equation}
Here $f^\mu_{(1)}$ and $f^\mu_{(2)}$ denote 
the first-order self-force and the second-order self-force, respectively.

To calculate the self-force we shall use a decomposition technique as a part 
of our derivation method (Several authors have 
previously used a similar technique 
together with other methods of derivation in calculating various types of self-forces,
 see for example \cite{Dirac,DB,DW}). 
Roughly speaking, in this technique one first decomposes the metric perturbations  
 into a certain regular piece and a certain singular piece. 
Next, one calculates the self-force and shows (using a certain method) 
 that the self-force is completely determined by the regular piece 
and does not depend on the singular piece. 
To apply such a technique to the derivation of  $f^\mu_{(1)}$ 
one might attempt to introduce the following decomposition 
 $h_{\mu\nu}=h^{sing}_{\mu\nu}+h^{reg}_{\mu\nu}$, 
where $h^{sing}_{\mu\nu}$ is a certain  singular piece and 
 $h^{reg}_{\mu\nu}$ is a certain regular piece. 
Here there is a difficulty, since this decomposition is nonunique. 
We can easily generate a family of such decompositions 
by invoking a transformation of the form 
 $h^{sing}_{\mu\nu}\rightarrow h^{sing}_{\mu\nu}+q_{\mu\nu}$, 
 $h^{reg}_{\mu\nu}\rightarrow h^{reg}_{\mu\nu}-q_{\mu\nu}$, where 
$q_{\mu\nu}$ is an arbitrary regular field. Therefore, there is a concern that
our final expression for  $f^\mu_{(1)}$  will depend 
on the specific decomposition in use. 
In this article we employ a specific decomposition (see below) that
suffers from a similar nonuniqueness problem. 
Nevertheless, we show that this nonuniqueness does
 not affect the formulas that 
we derive for   $f^\mu_{(1)}$ and $f^\mu_{(2)}$.

This article is organized as follows: 
In Sec. \ref{firsto} we derive an expression for the first-order gravitational 
self-force in a vacuum background spacetime. 
Tackling this simpler problem first allows us to introduce the 
main principles of our derivation method.
In addition, this analysis also serves as a check of our 
method, since we are able to compare our formula for $f^\mu_{(1)}$ with the previously derived 
expression for the first order self-force \cite{MST,QW}. 
Next, in Sec. \ref{secondo}  we generalize the principles which have
 been 
introduced in Sec. \ref{firsto}, and derive an expression for the second-order gravitational 
self-force in a vacuum background spacetime.

\section {First order self-force}\label {firsto}

We consider the first-order metric perturbations in the Lorenz gauge
\[
\bar{h}_{\mu\nu}^{\ \ \, ;\nu}=0\,,
\]
where overbar denotes the trace-reversal operator 
defined by
$\bar{h}_{\mu\nu}\equiv h_{\mu\nu} - (1/2)g_{\mu\nu}h_\alpha^{\  \alpha}$.
In the Lorenz gauge the first-order 
metric perturbations satisfy the following wave equation
\begin{equation}\label{waveeq}
\Box \bar{h}_{\mu\nu}+2R^{\eta\ \rho}_{\ \mu\ \nu}
\bar{h}_{\eta\rho}=-16\pi T_{\mu\nu}\,,
\end{equation}
where $\Box\equiv g^{\rho\sigma}\nabla_{\rho}\nabla_{\sigma}$, 
$R_{\eta\mu\rho\nu}$ denotes the Riemann tensor of the background geometry
with the sign convention of reference \cite{MTW}. Throughout this article we use 
the signature $(-,+,+,+)$ and geometrized units $G=c=1$.
In a local neighborhood of the worldline  
 the energy-momentum tensor (at the point particle limit) takes the form  of
\[
 T_{\mu\nu}(x)=\mu \int_{-\infty}^{\infty} \bar{g}^{\ \alpha}_{\mu}(x|z_G)u_{\alpha}
\bar{g}^{\ \beta}_{\nu}(x|z_G)u_{\beta}  \delta^4(x-z_G)
[-g(z_G)]^{-1/2} d\tau\,.
\]
where the four velocity is denoted $u^{\alpha}=\frac{dz_G^\alpha}{d\tau}$; 
for an arbitrary point $x'$ the notation 
$\bar{g}^{\ \alpha}_{\mu} =\bar{g}^{\ \alpha}_{\mu}(x|x')$ 
denotes the geodetic parallel propagator from $x'$ to $x$.
This bi-vector transports an arbitrary vector $A_{\alpha}(x')$ to a vector 
 $A_\mu(x)=\bar{g}^{\ \alpha}_{\mu}(x|x')A_{{\alpha}}(x')$
by a parallel propagation of the vector  $A_{{\alpha}}$ 
along the geodesic connecting $x'$ and $x$ (see Appendix A); 
 $g$ denotes the determinant 
of the background metric, and $\delta^4(x-x')$ denotes the four-dimensional 
(coordinate) Dirac delta-function.  

Eq. (\ref{foso}) implies that $f^\mu_{(1)}$ depends 
on the metric perturbations $h_{\mu\nu}$, 
where hereafter $h_{\mu\nu}$ shall denote the retarded solution of Eq. (\ref{waveeq}).
In a local neighborhood of the worldline it is useful 
 expand the perturbations $h_{\mu\nu}$ using the distance from 
the worldline as a small parameter. 
For definiteness let us consider an 
arbitrary point on the world line $\hat{z}=z_G(\hat{\tau})$, 
and expand $h_{\mu\nu}$ on $\Sigma(\hat{\tau})$  as follows
\begin{equation}\label{hdec}
h_{\alpha\beta}(x)=
\bar{g}^{\ \hat{\alpha}}_{\alpha} \bar{g}^{\ \hat{\beta}}_{\beta}
\Biglb[c^{(-1)}_{\hat{\alpha}\hat{\beta}} \varepsilon^{-1}+
c^{(0)}_{\hat{\alpha}\hat{\beta}}+
c^{(1)}_{\hat{\alpha}\hat{\beta}} \varepsilon
+O(\varepsilon^2)\Bigrb]\,.
\end{equation}
Here the overhat indices refer to $\hat{z}$, 
$\varepsilon$ denotes the length of the geodesic connecting 
 $\hat{z}$ and $x$. 
 The coefficients
 $\{ c^{(n)}_{\hat{\alpha}\hat{\beta}}\}$ 
transform as tensors under a coordinate transformation at  $\hat{z}$, and 
 transform as scalars under a coordinate transformation at $x$. 
By definition these coefficients are independent of $\varepsilon$. The 
explicit expressions for the coefficients  $\{ c^{(n)}_{\hat{\alpha}\hat{\beta}}\}$ 
are not required in this paper,
but we should point out  that some of the coefficients are  
 nonlocal (in particular $c^{(1)}_{\hat{\alpha}\hat{\beta}}$) i.e., 
they depend on the entire past history of the worldline (see e.g. \cite{PR}).
Notice that the perturbations $h_{\mu\nu}$ diverge as $\varepsilon^{-1}$
and therefore the calculation of the self-force $f^\mu_{(1)}$ 
requires one to introduce a regularization method. 

Our regularization method is based on a decomposition of 
$h_{\mu\nu}$ into the following tensor fields in a local neighborhood of  $z_G(\tau)$: 
\begin{equation}\label{hihsr}
h_{\mu\nu}= h^{I}_{\mu\nu}+ h^{SR}_{\mu\nu}\,.
\end{equation}
Here $h^{I}_{\mu\nu}$
denotes an {\em instantaneous} piece and 
$h^{SR}_{\mu\nu}$ denotes a {\em sufficiently regular}
piece. 
We define an instantaneous piece $ h^{I}_{\mu\nu}$ as a piece 
that admits a local expansion on $\Sigma({\hat{\tau}})$, reading
\begin{equation}\label{hidef}
h^I_{\alpha\beta}(x)=
\bar{g}^{\ \hat{\alpha}}_{\alpha} \bar{g}^{\ \hat{\beta}}_{\beta}
\Biglb[d^{(-1)}_{\hat{\alpha}\hat{\beta}} \varepsilon^{-1}+
d^{(0)}_{\hat{\alpha}\hat{\beta}}+
d^{(1)}_{\hat{\alpha}\hat{\beta}} \varepsilon
\Bigrb]\,,
\end{equation}
where the  expansion coefficients 
$\{d^{(-1)}_{\hat{\alpha}\hat{\beta}},d^{(0)}_{\hat{\alpha}\hat{\beta}},
d^{(1)}_{\hat{\alpha}\hat{\beta}}\}$ are constructed only from combinations of 
 quantities in following list: 
 the metric tensor  $g_{\hat{\mu}\hat{\nu}}$ and tensors constructed from the metric and its derivatives 
(e.g. Riemann tensor and its covariant derivatives), hereafter we shall refer to 
these tensors as background tensors;
the four velocity $u^{\hat{\mu}}$; the unit tangent vector
 $\nabla_{\hat{\alpha}}\varepsilon$ which is tangent 
to the geodesic connecting $\hat{z}$ and $x$ (see Eq. \ref{epsgradeps} in Appendix A), this 
quantity transforms as a vector at $\hat{z}$ and as a scalar at $x$; and numerical
coefficients.
We define a sufficiently regular piece $h^{SR}_{\mu\nu}$ as a piece whose first-order  
derivative $\nabla_{\alpha}h^{SR}_{\mu\nu}$ is continuous in a local neighborhood of $z_G$, 
 and its higher-order derivatives are not too singular  
(More precisely we demand that 
$\varepsilon^{n-1}\nabla_{\delta_n}...\nabla_{\delta_1} h^{SR}_{\mu\nu;\gamma}$, where 
$n\ge 1$, remains bounded as $x\rightarrow \hat{z}$.).

Roughly speaking, the instantaneous piece captures the singularity content  
of $h_{\mu\nu}$, and therefore the piece  $h^{SR}_{\mu\nu}$ is sufficiently regular. 
Another important property of  $h^I_{\alpha\beta}(x)$ is the ``instantaneous'' property:  
unlike $h_{\mu\nu}$, the instantaneous piece $h^I_{\alpha\beta}(x)$ on $\Sigma(\hat{\tau})$ 
 does not depend on the past history of $z_G$, it depends only on instantaneous quantities 
defined on $\Sigma({\hat{\tau}})$.
 
A priori it is not clear that $h_{\mu\nu}$ 
can be decomposed into an instantaneous piece and a sufficiently regular piece. 
For now let us suppose without proof that this  decomposition 
 exists. Later (in Sec. IIE) we shall provide a specific prescription for  its
 construction. 
Notice that decomposition  (\ref{hihsr})
is nonunique, and we are at liberty to introduce an alternative decomposition that
satisfies the same conditions, reading 
\begin{equation}\label{anotherdec}
h_{\mu\nu}= (h^{I}_{\mu\nu}+k_{\mu\nu})+ (h^{SR}_{\mu\nu}-k_{\mu\nu})\,,
\end{equation} 
where $k_{\mu\nu}$  is both sufficiently regular and instantaneous
(i.e. it satisfies all the conditions that these fields 
satisfy), but is otherwise   
 completely arbitrary. At this stage we do not provide a specific prescription for the construction of 
$h^{I}_{\mu\nu}$ and $h^{SR}_{\mu\nu}$, and 
our derivation will be based only on the general properties that
define this decomposition.

Substituting Eq. (\ref{foso}) into Eq. (\ref{eom}) and using 
Eq. (\ref{hihsr}) we find that at the first-order [ignoring $O(\mu^3)$ corrections]
 the equation of motion reads 
\begin{equation}\label{eom1}
\mu \frac{D^2 z^\mu}{D\tau^2}=\mu^2  f^\mu_{(1)}[g_{\mu\nu},
h^I_{\mu\nu}+h^{SR}_{\mu\nu},z_G(\tau)]\,,
\end{equation}
The left hand side of this equation is dimensionless, and therefore 
the vector $f^\mu_{(1)}$  must have 
dimensions of $(Length)^{-2}$. Notice  that the first-order 
self-force can have a complicated nonlinear dependence on the 
background metric. However, it must be linear in $h_{\mu\nu}$, since terms which  
 are nonlinear in $h_{\mu\nu}$  must also come with higher powers of $\mu$ 
and therefore belong to higher order corrections to $f^\mu$.
Due to this linearity in $h_{\mu\nu}$ we may group the terms
in the expression for  $f^\mu_{(1)}$ into a piece originating from
 $\{g_{\mu\nu},h^I_{\mu\nu},z_G(\tau)\}$ denoted $f^\mu_{(1)I}$, and 
 a piece originating from 
 $\{g_{\mu\nu},h^{SR}_{\mu\nu},z_G(\tau)\}$ denoted 
$f^\mu_{(1)SR}$. We formally write this decomposition as 
\begin{equation}\label{f1dec}
f^\mu_{(1)}[g_{\mu\nu},h^I_{\mu\nu}+h^{SR}_{\mu\nu},z_G(\tau)]=
f^\mu_{(1)I}[g_{\mu\nu},h^I_{\mu\nu},z_G(\tau)]+
f^\mu_{(1)SR}[g_{\mu\nu},h^{SR}_{\mu\nu},z_G(\tau)]\,.
\end{equation}
Here both $f^\mu_{(1)SR}$ and  $f^\mu_{(1)I}$
are well defined vector fields with dimensions of $(Length)^{-2}$.

\subsection {Instantaneous piece}

We now introduce the principles of our derivation method that 
is used extensively in this article.  
Let us focus on the instantaneous piece $f^\mu_{(1)I}[g_{\mu\nu},h^I_{\mu\nu},z_G(\tau)]$.
For definiteness, and without loss of generality, 
 we calculate $f^{\hat{\mu}}_{(1)I}$ (i.e.  $f^{{\mu}}_{(1)I}$ at $\hat{z}$). 
To derive an expression  for  $f^{\hat{\mu}}_{(1)I}$ we
first make a list of all the possible tensors (at $\hat{z}$) 
that may be included in the 
expression for  $f^{\hat{\mu}}_{(1)I}$. 
We refer to these tensors as the {\em tensorial constituents} of  $f^{\hat{\mu}}_{(1)I}$.
In practice, our list of tensorial constituents is constructed 
by studying the possible  tensors that can be constructed from the 
quantities $\{g_{\mu\nu},h^I_{\mu\nu},z_G(\tau)\}$. 
Next, we combine these  tensorial constituents to find  all the
possible vector expressions for $f^{\hat{\mu}}_{(1)I}$. 
In this method we first make a list of  individual tensors 
(without their combinations), and consider the combinations only at the next step,
when we construct the possible vector expressions.
Notice that these vector expressions  
must be well defined and must have dimensions of $(Length)^{-2}$.  

The explicit dependence of $f^{\hat{\mu}}_{(1)I}$ 
on $z_G(\tau)$ and $g_{\mu\nu}$ implies that 
the four velocity and 
the  background tensors must be 
included in the list of the  tensorial constituents of
$f^{\hat{\mu}}_{(1)I}$.   
Since $z_G(\tau)$ is a geodesic,  
higher order covariant time derivatives of the four-velocity
 vanish.  
The dependence of $f^{\hat{\mu}}_{(1)I}$ 
on $h^I_{\mu\nu}$ implies that tensors that can be constructed from $h^I_{\mu\nu}$
 should also be included in our list of tensorial constituents. 
All these tensors can be expressed using the quantities that are 
listed in Eq.  (\ref{hidef}). 
We now show that these tensors  do not extend our list of tensorial constituents.
From its definition $h^I_{\mu\nu}$ depends on both the four-velocity  
and the background tensors. These tensors have already been included in our list. 
In addition,  $h^I_{\mu\nu}$  also depends on  $\nabla_{\hat{\alpha}}\varepsilon$. 
However, the tangent vector  $\nabla_{\hat{\alpha}}\varepsilon$
is well defined only for a specific value of $x$ but it becomes multivalued   
as we change $x$. Since the self-force at $\hat{z}$ 
is independent of $x$, it can not depend on  $\nabla_{\hat{\alpha}}\varepsilon$. 
Moreover, we can not generate a new well-defined 
vector by taking the limit $\varepsilon\rightarrow 0$ 
of  $\nabla_{\hat{\alpha}}\varepsilon$. This limit is ill-defined since its value 
depends on the direction from which the limit is taken. 
We therefore conclude that $\nabla_{\hat{\alpha}}\varepsilon$ must 
be excluded from our list. A similar argument shows that   
 $\varepsilon$ must also be excluded from our list.
In addition, $h^I_{\mu\nu}$ also depends on the parallel propagator 
$\bar{g}^{\ \hat{\alpha}}_{\alpha}$.
Since $\bar{g}^{\ \hat{\alpha}}_{\alpha}$ 
is a bi-vector rather then a tensor, it 
can not be included in our list of tensorial constituents.
However, one can construct a well-defined tensor from 
$\bar{g}^{\ \hat{\alpha}}_{\alpha}$, 
for example by taking the limit  $\varepsilon\rightarrow 0$ 
of  $\bar{g}^{\ \hat{\alpha}}_{\alpha}$ (or by taking the limit  $\varepsilon\rightarrow 0$ 
of its covariant derivatives) 
and thereby construct a tensor at $\hat{z}$.
Notice, however,  that $\bar{g}^{\ \hat{\alpha}}_{\alpha}$ 
is completely determined from the background metric and the geodesic connecting 
$\hat{z}$ and $x$. Therefore, 
any well defined  tensor (independent of $x$) that is 
constructed from $\bar{g}^{\ \hat{\alpha}}_{\alpha}$ 
must be a background tensor. We therefore 
conclude that the tensors that are constructed from  $h^I_{\mu\nu}$  do
 not extend our list of  tensorial constituents.
Finally, we need to consider the possibility that the Levi-Civita tensor
might appear in our list. Notice that a term with an even number of Levi-Civita tensors
does not introduce any new tensor to our list, since these combinations
can be expressed using combinations of Kronecker deltas\footnote{Any two Levi-Civita tensors 
$\varepsilon^{\alpha\beta\gamma\delta}\varepsilon_{\mu\nu\rho\eta}$ can be expressed
as a sum over terms of the form
$\pm\delta_{\mu}^{\alpha}\delta_{\nu}^{\beta}\delta_{\rho}^{\gamma}\delta_{\eta}^{\delta}$
 with appropriate index permutations. This sum
is often expressed as a $4\times 4$ determinant of Kronecker deltas, see e.g. \cite {LL}.} . It is therefore sufficient to consider a single 
Levi-Civita tensor in our final expression for $f^{\hat{\mu}}_{(1)I}$.

Summarizing the above analysis, we have found that the expression for  
$f^{\hat{\mu}}_{(1)I}$ must be composed from 
the following tensorial constituents: the four velocity, the 
background tensors, and a single Levi-Civita tensor.
The fact that  $f^{\hat{\mu}}_{(1)I}$ has a dimension of  $(Length)^{-2}$
further constrains our list of tensorial constituents.
A background tensor with a dimension of  $(Length)^{-2}$ may contain only 
the following quantities: $g_{\hat{\mu}\hat{\nu}}$, $g_{\hat{\mu}\hat{\nu},\hat{\alpha}}$, and 
$g_{\hat{\mu}\hat{\nu},\hat{\alpha}\hat{\beta}}$.
Clearly a combination of $g_{\hat{\mu}\hat{\nu}}$ together with 
$g_{\hat{\mu}\hat{\nu},\hat{\alpha}}$  
can not produce such a tensor, since we may always employ a locally flat coordinate system  
(at $\hat{z}$) in which $g_{\hat{\mu}\hat{\nu},\hat{\alpha}}=0$. Consequently, 
a background tensor that is 
constructed from  $g_{\hat{\mu}\hat{\nu}}$, $g_{\hat{\mu}\hat{\nu},\hat{\alpha}}$ 
and  has a dimensionality of $(Length)^{-2}$ must vanish. 
Therefore, the desired tensor must be linear  
in   $g_{\hat{\mu}\hat{\nu},\hat{\alpha}\hat{\beta}}$ in addition to a combination of 
 $g_{\hat{\mu}\hat{\nu}}$ and $g_{\hat{\mu}\hat{\nu},\hat{\alpha}}$. 
Ref. \cite{Weinberg} provides a uniqueness theorem that ensures us that the only 
background tensor that satisfies all 
these requirements is   
 the Riemann tensor $R_{\hat{\alpha}\hat{\beta}\hat{\gamma}\hat{\delta}}$  (up to a combination with $g_{\hat{\mu}\hat{\nu}}$). 
Tensors that include higher-order derivatives of the metric, or tensors which 
are nonlinear in the second derivative of the metric 
[e.g. tensors of the form $\nabla_{\hat{\epsilon}} R_{\hat{\alpha}\hat{\beta}\hat{\gamma}\hat{\delta}}$, or tensors that are quadratic in the Riemann tensor]
  have a ``wrong'' dimensionality of  $(Length)^{-(2+n)}$ where $n>0$, and are therefore  
excluded from  our list\footnote{Notice that tensors with dimensions  of $(Length)^{n}$,  
where $n>0$ [e.g. $(R_{\hat{\alpha}\hat{\beta}\hat{\gamma}\hat{\delta}}R^{\hat{\alpha}\hat{\beta}\hat{\gamma}\hat{\delta}})^{-2}R_{\hat{\mu}\hat{\nu}\hat{\eta}\hat{\rho}}$] must be excluded since they give rise to terms that are ill defined in flat spacetime.}
. We conclude that each term in $f^{\hat{\mu}}_{(1)I}$ must be constructed from
a single Riemann tensor $R_{\hat{\alpha}\hat{\beta}\hat{\gamma}\hat{\delta}}$ and 
 may also include the following dimensionless tensors: 
$u^{\hat{\mu}}$, $g_{\hat{\mu}\hat{\nu}}$, 
and a single Levi-Civita tensor $\varepsilon^{\hat{\alpha}\hat{\beta}\hat{\gamma}\hat{\delta}}$.

We now show that in vacuum 
it is impossible to construct a vector from our list of tensorial constituents. 
First we consider vectors that are constructed from:  
 $R_{\hat{\alpha}\hat{\beta}\hat{\gamma}\hat{\delta}}$, 
$g_{\hat{\mu}\hat{\nu}}$, and  $u^{\hat{\mu}}$. We find that all these vectors 
vanish by virtue of  the 
symmetries  $R_{\hat{\alpha}\hat{\beta}(\hat{\gamma}\hat{\delta})}=0$, 
  $R_{(\hat{\alpha}\hat{\beta})\hat{\gamma}\hat{\delta}}=0$ and the fact that 
$R_{\hat{\alpha}\hat{\beta}}=0$. Considering next vector expressions
involving   the tensor $\varepsilon^{\hat{\mu}\hat{\nu}\hat{\rho}\hat{\eta}}$
 leads to the same vanishing result: 
An attempt to contract three or four indices of 
 $R_{\hat{\alpha}\hat{\beta}\hat{\gamma}\hat{\delta}}$ with   $\varepsilon^{\hat{\mu}\hat{\nu}\hat{\rho}\hat{\eta}}$ 
vanishes due to the identity  
$R^{\hat{\alpha}}_{\ [\hat{\beta}\hat{\gamma}\hat{\delta}]}=0$, contracting a pair of indices 
of 
 $R_{\hat{\alpha}\hat{\beta}\hat{\gamma}\hat{\delta}}$ with  
   $\varepsilon^{\hat{\mu}\hat{\nu}\hat{\rho}\hat{\eta}}$ 
produces a tensor with two pairs of indices (a pair from each tensor). 
Each of these pairs is antisymmetric, and therefore it is impossible to combine 
the resulting tensor  with the four velocity and construct a vector.  
The remaining possibilities vanish trivially.

Since one can not construct a non-vanishing 
vector from the above list of tensorial constituents we
are led to the conclusion that
\begin{equation}\label{f1inst}
f^{{\mu}}_{(1)I}=0\,.
\end{equation}
Ori have used a similar argument
to show that the expression for $f^{\mu}_{(1)}$ does not consists of any 
local terms \cite{Oripc}. In this article we shall extend this argument
in the analysis of $f^{\mu}_{(2)}$ (see below). After 
this analysis was completed we have learned that Anderson, Flanagan and  Ottewill
 have further extended and simplified Ori's argument 
in their calculation of a quasi-local expansion of the first-order self-force \cite{AFO}.

\subsection {Sufficiently regular piece}

Following the method of Sec. IIA we 
begin our analysis by studying the tensorial constituents of 
the sufficiently regular piece of the self-force 
$f^{\hat{{\mu}}}_{(1)SR}[g_{{\mu}{\nu}},h^{SR}_{{\mu}{\nu}},z_G(\tau)]$.
Recall that $f^{\hat{{\mu}}}_{(1)SR}$ must be 
a well-defined vector with a dimension of $(Length)^{-2}$.
The dependence of $f^{\hat{{\mu}}}_{(1)SR}$ on $g_{{\mu}{\nu}}$ and $z_G(\tau)$
 implies that  the background tensors and the four-velocity must be included 
in our list of the tensorial constituents of $f^{\hat{\mu}}_{(1)SR}$. 
However, in Sec. IIA we found that in vacuum these tensorial 
constituents (together with the Levi-Civita tensor) 
can not be combined to give a non-vanishing vector with a dimension  
of $(Length)^{-2}$. Therefore, all the terms
in the expression for $f^{\hat{\mu}}_{(1)SR}$ must contain  
an explicit (linear) dependence on $h^{SR}_{{\mu}{\nu}}$.

Since  $h^{SR}_{{\mu}{\nu}}$ is sufficiently regular it admits the following local 
expansion on $\Sigma(\hat{\tau})$ (see Appendix A)
\begin{equation}\label{hsrexp}
h^{SR}_{\alpha\beta}(x)=
\bar{g}^{\ \hat{\alpha}}_{\alpha} \bar{g}^{\ \hat{\beta}}_{\beta}
\Biglb[ h^{SR}_{\hat{\alpha}\hat{\beta}}-
h^{SR}_{\hat{\alpha}\hat{\beta};\hat{\gamma}}
\varepsilon\varepsilon^{;\hat{\gamma}}+O(\varepsilon^2)\Bigrb]\,.
\end{equation}
For illustration consider the simplest case of static particle in flat spacetime. 
In this case Eq. (\ref{hsrexp}) is reduced to a Taylor expansion: 
using Cartesian coordinates $x^a=(x^1,x^2,x^3)$  
we find that in this case we have 
$\bar{g}^{\ \hat{\alpha}}_{\alpha}=\delta^{\ \hat{\alpha}}_{\alpha}$,
$\varepsilon=\sqrt{(x^a-\hat{z}^a)(x^b-\hat{z}^b)\delta_{ab}}$, and
 $\varepsilon\varepsilon^{,\hat{c}}=-(x^c-\hat{z}^c)$.

We now examine the terms in Eq. (\ref{hsrexp}) that 
may appear in our list of the tensorial constituents of $f^{\hat{\mu}}_{(1)SR}$.
Following Sec. IIA we find that the quantities  $\bar{g}^{\ \hat{\beta}}_{\beta}$, 
$\varepsilon$, and $\varepsilon^{;\hat{\gamma}}$ that appear in 
Eq. (\ref{hsrexp}) are excluded from our list.
We are therefore left with the terms 
$h^{SR}_{\hat{\alpha}\hat{\beta}}$, $h^{SR}_{\hat{\alpha}\hat{\beta};\hat{\gamma}}$,  
and possibly higher-order derivatives (if they exist).
Recall that in our notation $h^{SR}_{\hat{\alpha}\hat{\beta}}$ has a dimension
of $(Length)^{-1}$  [see Eq. (\ref{deltag})]. Therefore,  higher-order derivatives 
 of the form $h^{SR}_{\hat{\mu}\hat{\nu};\hat{\gamma}\hat{\delta}_1...\hat{\delta}_n}$,  
$n\ge 1$,  must have dimensions of  $(Length)^{-2-n}$. 
By dimensionality, these terms must be 
excluded from our list \cite{nonreg}.
Consequently, from the various terms 
that are listed in Eq. (\ref{hsrexp}) only  the terms 
$h^{SR}_{\hat{\alpha}\hat{\beta}}$ and 
$h^{SR}_{\hat{\alpha}\hat{\beta};\hat{\gamma}}$ can 
appear in the expression for $f^{\hat{\mu}}_{(1)SR}$.
For example we can construct the following vectors 
$(g^{\hat{\alpha}\hat{\nu}}+u^{\hat{\alpha}}u^{\hat{\nu}})u^{\hat{\mu}}u^{\hat{\rho}}
h^{SR}_{\hat{\mu}\hat{\nu};{\hat{\rho}}}$ and 
$(g^{\hat{\alpha}\hat{\rho}}+u^{\hat{\alpha}}u^{\hat{\rho}})u^{\hat{\nu}}u^{\hat{\mu}}
h^{SR}_{\hat{\mu}\hat{\nu};{\hat{\rho}}}$.
Since the set of all possible vector expressions that may 
contribute to  $f^{\hat{\mu}}_{(1)SR}$ is not an empty set, 
we can not immediately implement the method of  Sec. IIA.  
To get around this difficulty we employ 
a gauge transformation. 
Previous studies have shown that the self-force is a gauge dependent quantity 
(see \cite{BOGAUGE} and also discussion in section IIC below). Therefore,  
 we may employ a gauge transformation  
to simplify our analysis.
Consider a regular gauge transformation of the form  
$x^\mu\rightarrow x^\mu - \mu \xi^\mu$.
We denote the first-order perturbations in the new gauge with $h^F_{\mu\nu}$.
In this gauge we have   
$h^F_{\mu\nu}=h^{I}_{\mu\nu}+h^{SR(F)}_{\mu\nu}$, where
$h^{SR(F)}_{\mu\nu}\equiv h^{SR}_{\mu\nu}+\xi_{\mu;\nu}+\xi_{\nu;\mu}$.
Here we have included the entire gauge transformation
in the definition of the new sufficiently regular piece 
$h^{SR(F)}_{\mu\nu}$, thus leaving $h^{I}_{\mu\nu}$ gauge invariant.
We demand that  $\xi_{\mu;\nu\rho}$ will be  continuous
 in a local neighborhood of  the worldline, and thereby we guarantee that $h^{SR(F)}_{\mu\nu}$
will satisfy the conditions of a sufficiently regular piece.
We shall refer to the new gauge as Fermi-gauge.
In Fermi gauge Eq. (\ref{f1dec}) reads
\begin{equation}\label{f1decfermi}
f^\mu_{F(1)}[g_{\mu\nu},h^I_{\mu\nu}+h^{SR(F)}_{\mu\nu},z_G(\tau)]=
f^\mu_{(1)I}[g_{\mu\nu},h^I_{\mu\nu},z_G(\tau)]+
f^\mu_{(1)SR(F)}[g_{\mu\nu},h^{SR(F)}_{\mu\nu},z_G(\tau)]\,,
\end{equation}
where $f^\mu_{F(1)}$ is the first-order self-force in Fermi gauge
and $f^\mu_{(1)SR(F)}$ is its corresponding sufficiently regular piece.
We impose the following gauge conditions along the world line $z_G(\tau)$
 \footnote{Fermi gauge was 
first introduced in Ref. \cite{eran1}. In this reference the gauge vector $\xi^\mu$ 
was defined using Eqs. (\ref{1fermigauge},\ref{2fermigauge}) with the replacement 
$h^{SR}_{\mu\nu} \rightarrow h^{R}_{\mu\nu}$,  where $h^{R}_{\mu\nu}$
is  Detweiler and Whiting regular 
piece \cite{DW}. These different definitions coincide once  
we express  $h^{SR}_{\mu\nu}$ in terms of 
$h^{R}_{\mu\nu}$ in Eq. (\ref{hsrhr}) below.}  
\begin{eqnarray}\label{1fermigauge}
&&\left[h^{SR(F)}_{\mu\nu}\right]_{z_G(\tau)}\equiv\biglb[h^{SR}_{\mu\nu}+\xi_{\mu;\nu}+\xi_{\nu;\mu}\bigrb]_{z_G(\tau)}=0\,,
\\\label{2fermigauge}
&& \left[h^{SR(F)}_{\mu\nu;\rho}\right]_{z_G(\tau)}\equiv\Biglb[\biglb(
h^{SR}_{\mu\nu}+\xi_{\mu;\nu}+\xi_{\nu;\mu}\bigrb)_{;\rho}\Bigrb]_{z_G(\tau)}=0\,.
\end{eqnarray}
In complete analogy with the Lorenz gauge, 
each term in $f^{\hat{\mu}}_{(1)SR(F)}$ (in Fermi gauge) must   
 explicitly depend on 
one of the first two coefficients in 
the expansion of $h^{SR(F)}_{{\mu}{\nu}}$ [similar to Eq. (\ref{hsrexp})]. 
Since these coefficients vanish all these terms
must be equal to zero. 
We therefore conclude that  
\begin{equation}\label{f1sr}
f^{{\mu}}_{(1)SR(F)}=0\,.
\end{equation}

We now briefly explain how to construct Fermi-gauge.
By contracting Eq. (\ref{2fermigauge}) with $u^\rho$,  
we find that  $\frac{D}{D\tau}h^{SR(F)}_{\mu\nu}=0$. 
This result is consistent with Eq. (\ref{1fermigauge}).
We can now solve Eq. (\ref{1fermigauge}) at an initial point $z_G(\tau_0)$.
Satisfying Eq. (\ref{2fermigauge}) guarantees the validity 
of  Eq. (\ref{1fermigauge}) everywhere along $z_G(\tau)$. 
We construct an arbitrary gauge vector $\xi_{(0)\mu}$ 
at $z_G(\tau_0)$, and demand that
\begin{equation}\label{initcon}
\xi_{(0)\mu;\nu}=\xi_{(0)\nu;\mu}=-\frac{1}{2} h^{SR}_{\mu\nu}(\tau_0)\,,
\end{equation}
thereby satisfying Eq. (\ref{1fermigauge}).  
We now turn to Eq. (\ref{2fermigauge}). 
Viewing these equations as algebraic equations,
we find that we have a set of $40$
equations for the $64$ variables $\xi_{\mu;\nu \rho}$.
By introducing  the commutation relation 
\begin{equation}\label{commute}
2\xi_{\mu;[\nu\alpha]}=R^{\epsilon}_{\ \mu\nu\alpha}\xi_{\epsilon}\,,
\end{equation}
we include another $24$ equations which brings us to $64$ equations, as desired.
Using the identities of the Riemann tensor we obtain from
 Eqs. (\ref{2fermigauge},\ref{commute}) the following relation 
\begin{equation}\label{gradgradxi}
\xi_{\nu;\mu\alpha}=R^{\epsilon}_{\ \alpha\mu\nu}\xi_{\epsilon}
-\delta \Gamma_{\nu\mu\alpha}^{SR}\,,
\end{equation}
where
\[
\delta \Gamma_{\nu\mu\alpha}^{SR}=\frac{1}{2}(h^{SR}_{\mu\nu;\alpha}+h^{SR}_{\nu\alpha;\mu}-
h^{SR}_{\mu\alpha;\nu})
\]
Here all quantities are evaluated on the worldline.
By contracting Eq. (\ref{gradgradxi}) with $u^\alpha u^\mu$
we obtain 
\begin{equation}\label{transxi}
\ddot{\xi}_\nu=R^{\epsilon}_{\ \alpha\mu\nu}\xi_{\epsilon}u^\alpha u^\mu-
u^\alpha u^\mu \delta \Gamma_{\nu\mu\alpha}^{SR}\,,
\end{equation}
where $\ddot{\xi}_{\nu}\equiv\frac{D^2\xi_{\nu}}{D\tau^2}$.
One can solve this second-order transport equation 
by using $\xi_{(0)\mu}$, and $\dot{\xi}_{(0)\mu}$ 
as initial conditions, and thereby obtain a 
gauge vector $\xi_{\mu}(\tau)$  along the worldline. 
Similarly we can construct a first-order transport equation for $\xi_{\mu;\nu}$ 
by contracting Eq. (\ref{gradgradxi}) with $u^\alpha$. 
Using  $\xi_{\mu}(\tau)$  together with Eq. (\ref{initcon}) this first-order transport 
equation can be integrated to give $\xi_{\mu;\nu}(\tau)$. $\xi_{\mu;\nu\alpha}(\tau)$
is then obtained by substituting $\xi_{\mu}(\tau)$ into Eq. (\ref{gradgradxi}).
This completes the formal solution of  Eqs. (\ref{1fermigauge},\ref{2fermigauge}), and allows one
to calculate the leading terms in 
a local expansion of $\xi_\mu$ in the vicinity of the worldline. This local 
expansion could be arbitrarily continued, thereby obtaining a global 
 definition for Fermi gauge.

\subsection {First-order self-force in Fermi gauge}

Combining Eqs. (\ref{f1inst},\ref{f1sr}) together with Eq. (\ref{f1decfermi}) 
we find that in Fermi gauge the first-order self-force vanishes, namely
\begin{equation}\label{f1fermi}
f^{\mu}_{F(1)}=0\,.
\end{equation}
This result might appear surprising at first. 
The surprise is removed once one recalls that within the framework of
general relativity  
the complete information about the motion of a point like object 
includes two pieces of information: the object's worldline 
together with the spacetime metric.
 It is therefore not surprising that 
one may transfer information about the motion of the object from the description
of the worldline to the description of the gauge. This also implies that   
an expression for the self-force is physically meaningful only if 
the gauge is specified.

\subsection {First-order self-force in the Lorenz gauge}

We now derive the expression for the first-order self-force in the Lorenz gauge.
For this purpose, we consider the inverse gauge transformation i.e. a transformation 
from Fermi gauge to the original Lorenz gauge, generated by 
$x^\alpha \rightarrow x^\alpha+\mu\xi^\alpha$, where $\xi^\alpha$
 satisfies Eqs. (\ref{1fermigauge},\ref{2fermigauge}). 
Barack and Ori have derived the gauge transformation formula for the first-order self-force 
\cite{BOGAUGE}.
Their analysis shows that under a regular gauge transformation of the form 
$x^\alpha\rightarrow x^\alpha - \mu\xi^\alpha$ the first-order self-force 
transforms as 
\begin{equation}\label{ftrans}
f^{\alpha}_{(1)}\rightarrow f^{\alpha}_{(1)}+\delta f^{\alpha}_{(1)}\,,
\end{equation}
where  
\begin{equation}\label{deltaf1}
\delta f^{\alpha}_{(1)}=
-(g^{\alpha\lambda}+u^{\alpha}u^{\lambda})\ddot{\xi}_{\lambda}-
R^{\alpha}_{\ \mu\lambda\nu}u^{\mu}\xi^{\lambda}u^{\nu}\,.
\end{equation}
Using Eqs. 
(\ref{f1fermi},\ref{ftrans},\ref{deltaf1}) we find that 
in the Lorenz gauge the self-force is given by
\begin{equation}\label{f1deltaf1}
 f^{\alpha}_{(1)}= f^{\alpha}_{F(1)}-\delta f^{\alpha}_{(1)}=
-\delta f^{\alpha}_{(1)}\,,
\end{equation}
where the minus sign originate from the fact that we are considering
the inverse gauge transformation. 
We now calculate $f^{\alpha}_{(1)}$ by first 
substituting Eq. (\ref{transxi}) into Eq. (\ref{deltaf1}), and then substituting
 $\delta f^{\alpha}_{(1)}$ into Eq. (\ref{f1deltaf1}). In this way 
we obtain the following expression for 
the first-order self-force in the Lorenz gauge
\begin{equation}\label{f1hsr}
\mu^2 f^{\mu}_{(1)}=\mu^2 K^{\mu\alpha\beta\gamma} h^{SR}_{\alpha\beta;\gamma}\,.
\end{equation}
where 
\[
K^{\mu\alpha\beta\gamma}=-g^{\mu\beta}u^{\gamma}u^{\alpha}
-(1/2) u^\mu u^\alpha u^\beta u^\gamma + (1/2)g^{\mu\gamma}u^{\alpha}u^{\beta}
\,.
\]

We have already mentioned that decomposition (\ref{hihsr}) is nonunique, and one
 can generate a family of alternative decompositions (\ref{anotherdec}),  
where $k_{\mu\nu}$  is both sufficiently regular and instantaneous, 
but is otherwise completely arbitrary. 
Clearly our result in Eq. (\ref{f1hsr}) is self-consistent 
only if it is invariant under the transformation 
$h^{SR}_{\alpha\beta}\rightarrow h^{SR}_{\alpha\beta}-k_{\alpha\beta}$. This invariance follows from  properties of 
$k_{\alpha\beta}$. Since $k_{\alpha\beta}$ is sufficiently regular 
the vector  $K^{\mu\alpha\beta\gamma} k_{\alpha\beta;\gamma}$ is well defined
on the worldline, and since $k_{\alpha\beta}$ is instantaneous it follows from 
 subsection IIA that  
$K^{\mu\alpha\beta\gamma} k_{\alpha\beta;\gamma}=0$, meaning that Eq. (\ref{f1hsr}) is 
invariant under the transformation $h^{SR}_{\alpha\beta}\rightarrow h^{SR}_{\alpha\beta}-k_{\alpha\beta}$.

\subsection {Specific construction of the metric decomposition}

So far we did not specify how to practically construct  decomposition (\ref{hihsr}), 
nor did we show that this decomposition exists.
In this subsection we complete our derivation by  
providing a specific prescription for the construction of decomposition (\ref{hihsr}).
Decomposition  (\ref{hihsr}) can be constructed in more than one method. 
For simplicity  we choose to follow a previous  
 analysis by  Detweiler and  Whiting \cite{DW} and relate 
 decomposition (\ref{hihsr}) to their decomposition, reading
\begin{equation}\label{dw}
h_{\alpha\beta}=h_{\alpha\beta}^S+h_{\alpha\beta}^R\,.
\end{equation}
Here the (trace reversed) 
singular piece  $\bar{h}_{\alpha\beta}^S$ satisfies the inhomogeneous wave-equation 
(\ref{waveeq}). In a local neighborhood of the worldline it is defined by 
\begin{equation}\label{hs}
\bar{h}^S_{\mu\nu}(x)
=4\int_{-\infty}^{\infty}
 G_{\mu\nu\alpha\beta}^{S}[x|z(\tau)]u^\alpha(\tau)u^\beta(\tau) d\tau\,,
\end{equation} 
where the singular Green's function is given by
\begin{equation}\label{gs}
G_{\mu\nu\alpha'\beta'}^{S}[x|x']=
\frac{1}{2}\Biglb[U_{\mu\nu\alpha'\beta'}[x|x']\delta(\sigma)+
V_{\mu\nu\alpha'\beta'}[x|x']\theta(-\sigma)
\Bigrb]\,.
\end{equation}
Here $\sigma=\sigma(x|x')$ denotes  half the square of the invariant
distance measured along a geodesic connecting $x$ and $x'$; $U_{\mu\nu\alpha'\beta'}[x|x']$ and
 $V_{\mu\nu\alpha'\beta'}[x|x']$ are certain regular bi-tensors (for their
definitions and properties see e.g. \cite{Friedlander,PR}),
$\theta$ denotes a step function. Eq. (\ref{gs}) implies that for 
a fixed $x$ only a finite interval
contributes to the integral in 
Eq. (\ref{hs}). This interval approaches zero as 
as the evaluation point $x$ approaches point $\hat{z}$ on the worldline.
The regular field $\bar{h}_{\alpha\beta}^R$ satisfies a homogeneous wave equation 
[Eq. (\ref{waveeq}) with $T_{\mu\nu}=0$], and is a smooth field in the vicinity 
of  the worldline.

Expanding  $\bar{h}^S_{\mu\nu}$ on the hypersurface of constant time 
 $\Sigma(\hat{\tau})$ gives (see Appendix B)
\begin{equation}\label{hsexpand}
\bar{h}_{\alpha\beta}^S(x)=
\bar{g}_{(\alpha}^{\ \ \hat{\alpha}} \bar{g}_{\beta)}^{\ \hat{\beta}}
\Biglb[4u_{\hat{\alpha}}u_{\hat{\beta}} \varepsilon^{-1}
-\biglb(
\frac{2}{3}R_{\hat{\mu}\hat{\gamma}\hat{\nu}\hat{\delta}}
u^{\hat{\mu}} u^{\hat{\nu}} \varepsilon^{;\hat{\gamma}}
\varepsilon^{;\hat{\delta}}
u_{\hat{\alpha}} u_{\hat{\beta}}+
4 R_{\hat{\mu}\hat{\alpha} \hat{\nu}\hat{\beta}}
u^{\hat{\mu}} u^{\hat{\nu}}\bigrb)\varepsilon
+O(\varepsilon^2)\Bigrb]\,.
\end{equation}
Comparing Eq. (\ref{hsexpand}) with Eq. (\ref{hidef})
we find that the leading terms in the
 expansion   $\bar{h}_{\alpha\beta}^S(x)$
[and ${h}_{\alpha\beta}^S(x)$] satisfy the conditions
of an instantaneous piece. We therefore identify  $\bar{h}_{\alpha\beta}^I(x)$ 
with the expression in  Eq. (\ref{hsexpand}) 
up to $O(\varepsilon)$ inclusive. This identification implies that 
in the vicinity of the worldline we have
\begin{equation}\label{hsrhr}
h_{\alpha\beta}^{SR}=h_{\alpha\beta}^{R}+O(\varepsilon^2)\,,
\end{equation}
where the $O(\varepsilon^2)$ discrepancy between $\bar{h}_{\alpha\beta}^{SR}$
and $h_{\alpha\beta}^{R}$
 originate from terms of order  $\varepsilon^2$
 which appear in $\bar{h}_{\alpha\beta}^S$ but are absent from 
 $\bar{h}_{\alpha\beta}^I$. Eq. (\ref{hsrhr}) implies 
that the expression that have been identified as $h_{\alpha\beta}^{SR}$ satisfies
the required condition  of a sufficiently regular piece\footnote{Notice that the singularities 
of the derivatives of the $O(\varepsilon^2)$ term in Eq. (\ref{hsrhr}) 
are constrained by the form of the  $O(\varepsilon^2)$  terms in 
Eq. (\ref{generalform}) in Appendix B, as required.}. 
We therefore conclude that we have constructed a decomposition 
which satisfies the conditions of decomposition (\ref{hihsr}).
Substituting Eq. (\ref{hsrhr}) into Eq. (\ref{f1hsr}) yields
the standard expression for the first-order self-force \cite{DW}
\begin{equation}\label{f1hr}
\mu^2 f^{\mu}_{(1)}=\mu^2 K^{\mu\alpha\beta\gamma} h^{R}_{\alpha\beta;\gamma}\,.
\end{equation}

\subsection {Summary}

Before we generalize our method and tackle the second-order problem it 
is instructive
to summarize the derivation method that we have used.
First, in Eq. (\ref{eom}) we introduced  the general form of the 
equation of motion. Next,  in Eq. (\ref{hihsr})
we introduced a decomposition of the first-order metric perturbations
into instantaneous and sufficiently regular pieces, 
and defined their properties. Using these 
properties we showed in Sec. IIA that one can not construct  
a well defined vector (with the desired dimensionality and the correct scaling in $\mu$) 
from the tensorial constituents of the instantaneous piece. 
Next, in Sec. IIB, we invoked a gauge transformation from
 Lorenz-gauge to Fermi-gauge, and used the same method (as in  Sec. IIA)
 to show that the sufficiently regular piece of the first-order self-force vanishes. 
At this stage we concluded that in Fermi-gauge the first-order self-force
 must  vanish. 
Next, in Sec. IID, we used the inverse gauge transformation together with Barack and Ori 
transformation formula to derive the expression for the first-order self-force 
in the original Lorenz gauge.
Finally, in Sec. IIE, we employed Detweiler and Whiting decomposition and provided 
a prescription for the construction of decomposition (\ref{hihsr}).
Using this prescription we have obtained the standard expression 
for the first-order self-force in the Lorenz gauge, given by Eq. (\ref{f1hr}) above.
The fact that the entire derivation followed from the general 
properties of decomposition (\ref{hihsr})  led to a  simple analysis.  
This simplicity enables us now to tackle the more complicated second-order problem. 

\section {Second order self-force}\label{secondo}

From its definition the second-order 
self-force depends on the second-order metric perturbations.
It is therefore necessary to begin our discussion by introducing
a construction method for these second-order perturbations. 
Already this preliminary stage introduces difficulties.
The difficulties and their resolution were studied in 
Refs. \cite{eran1,eran2}. We now briefly summarize the main
results of these references.

\subsection {Construction of the second-order metric perturbations}

Following Refs. \cite{eran1,eran2} 
 we specialize to a compact object which is a Schwarzschild black-hole with a mass $\mu$.
We focus on the region far from the black hole i.e.
 at distances larger than $r_E$, where $r_E(\mu)\gg\mu$, and refer
to this region as the external zone.
In the external zone 
the geometry is dominated by the background geometry, and the full metric is 
represented  by Eq. (\ref{fullg}), where the metric perturbations are given by Eq.  
 (\ref{deltag}). We let $r_E(\mu)\rightarrow 0$ as $\mu \rightarrow 0$. 
In this limit we write the perturbations equations as
\begin{eqnarray}\label{h1eq}
&&D_{\mu\nu}[\bar{h}]=0\ ,\ x\not\in \gamma\,, \\\label{h2eq}
&&D_{\mu\nu}[\bar{l}]=S_{\mu\nu}[\bar{h}]\ ,\ x\not\in \gamma\,.
\end{eqnarray}
Here $\gamma$ is a timelike worldline, the operators 
$D_{\mu\nu}$ and $S_{\mu\nu}$ are obtained from an expansion of the
 full Ricci tensor, where 
for brevity we have omitted tensorial indices inside the squared brackets. 
This expansion is obtained by substituting  Eq. (\ref{fullg})
into the Ricci tensor, which gives $R^{full}_{\mu\nu}=R_{\mu\nu}^{(0)}+R_{\mu\nu}^{(1)}[\delta g]+
R_{\mu\nu}^{(2)}[\delta g]+O(\delta g^3)$ (For explicit expressions 
of the terms in this expansion see e.g. \cite{MTW}.).  
To simplify the notation we denote $\bar{R}_{\mu\nu}^{(1)}[h]$ with $D_{\mu\nu}[\bar{h}]$, 
and denote $-\bar{R}_{\mu\nu}^{(2)}[h]$ with $S_{\mu\nu}[\bar{h}]$, where
$h_{\mu\nu}$ is expressed in terms of $\bar{h}_{\mu\nu}$.
D'Eath have shown that in the limit, the perturbations $h_{\mu\nu}$  
are identical to the retarded first-order metric-perturbations  
produced by a unit-mass point particle tracing  
the worldline $\gamma$  \cite{DEATH}. At the leading order of approximation, $\gamma$
 is a geodesic of the background spacetime \cite{DEATH}, which we continue to denote
with  $z_G(\tau)$. Therefore, in the Lorenz gauge the perturbations $h_{\mu\nu}$  
are given  by the retarded solution of Eq. (\ref{waveeq}).
At the next order of approximation, the worldline $\gamma$ has an $O(\mu)$ 
acceleration, induced by the first order self-force. 
Hereafter we shall adopt Fermi gauge for the first-order perturbations.
Recall that in this gauge the first-order self-force vanishes  and
the geodesic worldline $z_G(\tau)$ becomes sufficiently accurate up to 
acceleration of order $O(\mu)$, inclusive. In this gauge we may replace $\gamma$ 
 with $z_G(\tau)$ when we construct the second-order metric perturbations \cite{eran1,eran2}. 

Refs. \cite{eran1,eran2} show that the second-order metric perturbations 
can be decomposed as follows  
\begin{equation}\label{final}
\bar{l}_{\mu\nu}=\bar{\psi}_{\mu\nu}+\delta\bar{l}_{\mu\nu}\,.
\end{equation}
where the first piece $\bar{\psi}_{\mu\nu}$  is given by the following 
 linear combination of 
terms that are quadratic in $\bar{h}^F_{\mu\nu}$
\begin{equation}\label{psiexplicit}
\bar{\psi}_{\mu\nu}=\frac{1}{64}\left[2(c_A\varphi^{A}_{\mu\nu}+
c_B\varphi^{B}_{\mu\nu})-7(c_C\varphi^{C}_{\mu\nu}+
c_D\varphi^{D}_{\mu\nu})\right]\,. 
\end{equation}
Here 
$\varphi^{A}_{\mu\nu}=\bar{h}^{F\rho}_{\ \ \ \mu}
\bar{h}^{F}_{\rho\nu}$, $\varphi^{B}_{\mu\nu}=\bar{h}^{F\rho}_{\ \ \ \rho}
\bar{h}^{F\ \ \ \ }_{\mu\nu}$, $\varphi^{C}_{\mu\nu}=(\bar{h}^{F\eta\rho}
\bar{h}^{F}_{\ \ \eta\rho})g_{\mu\nu}$,  
$\varphi^{D}_{\mu\nu}=\left(\bar{h}^{F\rho}_{\ \ \ \rho}\right)^2
g_{\mu\nu}$; and  the constants $c_A,c_B,c_C,c_D$ must satisfy
$c_A+c_B=1$, $c_C+c_D=1$, 
but are otherwise arbitrary.
The second piece  $\delta\bar{l}_{\mu\nu}$ is given by the retarded
solution of the following wave equation  
\begin{equation}\label{deltah2final}
\Box \delta\bar{l}_{\mu\nu}+2R^{\eta\ \rho}_{\ \mu\ \nu}
\delta\bar{l}_{\eta\rho}=-2\delta S_{\mu\nu} \,.
\end{equation}
Here $\delta S_{\mu\nu}\equiv S_{\mu\nu}-D_{\mu\nu}[\bar{\psi}]$.

\subsection {General considerations}

The second term in Eq. (\ref{foso}),  namely
\begin{equation}\label{so}
\mu^3 f^\mu_{(2)}[g_{\mu\nu},h^F_{\mu\nu},l_{\mu\nu},z_G(\tau)]
\end{equation}
provides a complete list of  the quantities that the second order self-force may depend upon.
Notice that  working in Fermi gauge justifies the substitution  
$z(\tau)\rightarrow z_G(\tau)$ which was made in Eq. (\ref{so}).
This equation also states that the second-order self-force 
is proportional to $\mu^3$, meaning  that this self-force  induces an acceleration which scales as $\mu^2$.
To obtain the desired $\mu^2$ scaling of the acceleration, the second-order self-force
must be linear in  $\mu^2 l_{\mu\nu}$. 
Higher powers of  $\mu^2 l_{\mu\nu}$ are therefore excluded from $f^\mu_{(2)}$. 
Similarly, terms in  $f^\mu_{(2)}$ that depend 
explicitly on $h^F_{\mu\nu}$  must be quadratic in  $h^F_{\mu\nu}$.
Combining all the quantities  that  $f^\mu_{(2)}$ depends upon
together with Eq. (\ref{final}) we obtain 
\begin{equation}\label{f2ab}
f_{(2)}^\mu=f^{\mu}_{(2)A}[g_{\alpha\beta},z_G(\tau),h^F_{\gamma\delta}]
+f^{\mu}_{(2)B}
[g_{\alpha\beta},z_G(\tau),\delta l_{\gamma\delta}]\,.
\end{equation}
Here the first term is quadratic in $h^F_{\mu\nu}$ and the second term is linear in 
 $\delta l_{\gamma\delta}$. 

From Eqs. (\ref{eom},\ref{foso}) we find that 
 $f^\mu_{(2)}$ has a dimension of  $(Length)^{-3}$.
Following Sec. II, we shall list all the tensorial constituents of 
 $f^{\mu}_{(2)}$. The dimensionality of  $f^{\mu}_{(2)}$ restricts 
 our list of tensorial constituents to tensors with
 dimensionality of  $(Length)^{-n}$, $0\le n\le3$. 
This follows from the fact that we do not 
have tensors with dimensionality of  $(Length)^{j}$, $j>0$ at our disposal.

\subsection{First term $f^{\mu}_{(2)A}$}

First we focus on the term  
$f^{\mu}_{(2)A}[g_{\alpha\beta},z_G(\tau),h^F_{\gamma\delta}]$ in Eq. (\ref{f2ab}). 
Here it proves simpler to consider a general
dependence  (not necessarily quadratic)  of $f^{\mu}_{(2)A}$
on  $h^F_{\mu\nu}$. In the following analysis we shall often use   
 decomposition (\ref{dw}). 
We write this decomposition in Fermi gauge, and obtain 
\begin{equation}\label{dwfermi}
h^F_{\alpha\beta}=h_{\alpha\beta}^S+h_{\alpha\beta}^{R(F)}\,,\,
 h^{R(F)}_{\alpha\beta}\equiv h^{R}_{\alpha\beta}+\xi_{\alpha;\beta}+\xi_{\beta;\alpha}\,,
\end{equation}
where $\xi_{\alpha}$ satisfies Eqs. (\ref{1fermigauge},\ref{2fermigauge}).
Using this decomposition we schematically write 
the first piece as 
\begin{equation}
f^{\mu}_{(2)A}[g_{\alpha\beta},z_G(\tau),h^F_{\gamma\delta}]=
f^{\mu}_{(2)A}[g_{\alpha\beta},z_G(\tau),h^S_{\gamma\delta}+h^{R(F)}_{\gamma\delta}]\,.
\end{equation}
We shall now list all   the 
 tensorial constituents of $f^{\mu}_{(2)A}$.
First, we shall consider  the quantities  $\{g_{\alpha\beta},z_G(\tau),h^S_{\gamma\delta}\}$,
 list the relevant tensors, and discuss
the possible expressions for $f^{\mu}_{(2)A}$ that can be constructed from 
these quantities . Next, we shall consider all the remaining possibilities.

\subsubsection{$h^S$  terms}

We now list all the tensorial constituents  
 that can be constructed from the quantities 
$\{g_{\alpha\beta},z_G(\tau),h^S_{\gamma\delta}\}$ and have dimensions  of 
 $(Length)^{-n}$, $0\le n\le3$.  
From the quantities  $\{g_{\alpha\beta},z_G(\tau)\}$ we obtain the following list:   
$u^{\hat{\mu}}$, $g_{\hat{\mu}\hat{\nu}}$,
 $R_{\hat{\alpha}\hat{\beta}\hat{\gamma}\hat{\delta}}$,
 and  $R_{\hat{\alpha}\hat{\beta}\hat{\gamma}\hat{\delta};\hat{\epsilon}}$. 
In addition, our list includes the Levi-Civita tensor. 
Recall that $\bar{h}^S_{\gamma\delta}$ 
admits the local expansion (\ref{hsexpand}). The coefficients of this expansion    
 up to $O(\varepsilon)$ (inclusive) satisfy the conditions of an instantaneous piece. 
More generally, the higher order coefficients in this expansion 
are composed from the same kind of instantaneous quantities, namely: background tensors,
 $u^{\hat{\mu}}$, $\nabla_{\hat{\alpha}}\varepsilon$, numerical
coefficients (see Appendix B). 
Following the arguments of Sec. IIA we now find that $h^S_{\gamma\delta}$ 
 does not add any new tensor to the above list of tensorial constituents. 
Examining this list we find that a vector with the desired dimensionality of 
 $(Length)^{-3}$ 
must be linear in  $R_{\hat{\alpha}\hat{\beta}\hat{\gamma}\hat{\delta};\hat{\epsilon}}$ 
and may also include combinations of 
the following dimensionless 
tensors: $u^{\hat{\mu}}$, $g_{\hat{\mu}\hat{\nu}}$ and  $\varepsilon^{\hat{\alpha}\hat{\beta}\hat{\gamma}\hat{\delta}}$.

Examining our list of tensorial constituents shows 
that they can not be combined to give a vector with a dimension
of $(Length)^{-3}$:
Consider first vectors  that are constructed 
from the following tensors $R_{\hat{\alpha}\hat{\beta}\hat{\gamma}\hat{\delta};\hat{\epsilon}}$, $u^{\hat{\mu}}$, and $g_{\hat{\mu}\hat{\nu}}$.
By virtue of  the symmetries  $R_{\hat{\alpha}\hat{\beta}(\hat{\gamma}\hat{\delta})}=0$, 
  $R_{(\hat{\alpha}\hat{\beta})\hat{\gamma}\hat{\delta}}=0$ and the fact that 
$R_{\hat{\alpha}\hat{\beta}}=0$, we find that up to a sign the only
possibility is   $R^{\hat{\epsilon}}_{\ \hat{\beta}\hat{\gamma}\hat{\delta};\hat{\epsilon}}
u^{\hat{\beta}}u^{\hat{\delta}}$. This possibility vanishes by virtue of 
the  Bianchi identities which imply that in vacuum we have  
$R^{\hat{\epsilon}}_{\ \hat{\beta}\hat{\gamma}\hat{\delta};\hat{\epsilon}}=0$. 
Similarly, using the properties of the Riemann tensor together with properties of the Levi-Civita tensor 
one finds that all the  vector expressions composed from the tensor 
$\varepsilon^{\hat{\alpha}\hat{\beta}\hat{\gamma}\hat{\delta}}$ 
together with  $R^{\hat{\epsilon}}_{\ \hat{\beta}\hat{\gamma}\hat{\delta};\hat{\epsilon}}$, 
$g_{\hat{\mu}\hat{\nu}}$, and  $u^{\hat{\mu}}$, vanish as well. 
We conclude that by combining tensorial constituents that originate from 
the list   $\{g_{\alpha\beta},u^\mu,h^S_{\gamma\delta}\}$  
one can not construct a vector with  dimensionality
of $(Length)^{-3}$.

\subsubsection{Remaining terms}

Since the tensors  $\{g_{\alpha\beta},u^\mu,h^S_{\gamma\delta}\}$ can not produce
a vector with the desired dimensionality  we find that each term in $f^{\mu}_{(2)A}$ 
must  explicitly depend on $h^{R(F)}_{\alpha\beta}$.
Expanding  $h^{R(F)}_{\alpha\beta}$ in a local neighborhood of the worldline 
gives
\begin{equation}\label{hrfexp}
h^{R(F)}_{\alpha\beta}(x)=\bar{g}_{\alpha}^{\ \hat{\alpha}}
\bar{g}_{\beta}^{\ \hat{\beta}}
\left[\frac{1}{2}\varepsilon^2 
h^{R(F)}_{\hat{\alpha}\hat{\beta};\hat{\gamma}\hat{\delta}}\varepsilon^{;\hat{\gamma}}
\varepsilon^{;\hat{\delta}}
+O(\varepsilon ^3)\right]\,.
\end{equation}
Notice that the leading terms that 
scale as $\varepsilon^0$ and  $\varepsilon^1$ vanished 
by virtue of Fermi gauge.  
In this expansion only the coefficient of the $\varepsilon^2$ term has a  
dimension which fits our dimensionality condition of 
 $(Length)^{-n}$, $0\le n\le3$. Therefore, only this 
coefficient can contribute to our list of tensorial constituents. 
To construct a vector with a dimension  of  $(Length)^{-3}$
we must combine   
$h^{R(F)}_{\hat{\alpha}\hat{\beta};\hat{\gamma}\hat{\delta}}$ with the other 
dimensionless tensors from our list, namely $u^{\hat{\mu}}$, $g_{\hat{\mu}\hat{\nu}}$, and
$\varepsilon^{\hat{\alpha}\hat{\beta}\hat{\gamma}\hat{\delta}}$.
One can easily construct such a vector, for example 
$(g^{\hat{\alpha}\hat{\nu}}+u^{\hat{\alpha}}u^{\hat{\nu}})
u^{\hat{\mu}}u^{\hat{\rho}}u^{\hat{\epsilon}}
h^{R(F)}_{\hat{\nu}\hat{\mu};\hat{\rho}\hat{\epsilon}}$. Following Sec. IIB
we  use the gauge freedom to eliminate all the   
possibilities of constructing such a vector. We extend the definition of 
Fermi gauge by restricting the values of  
$\xi_{{\mu};{\nu}{\alpha}{\beta}}(\tau)$.
From Eq. (\ref{dwfermi}) we obtain
\begin{equation}\label{gradgradhr}
 h^{R(F)}_{\mu\nu;\alpha\beta}
=
h^{R}_{\mu\nu;\alpha\beta}+\xi_{\mu;\nu\alpha\beta}+
\xi_{\nu;\mu\alpha\beta}\,.
\end{equation}
Following  Eq.(\ref{gradgradxi}) we introduce  the anzats
\begin{equation}\label{gradgradgradxi}
\xi_{{\nu};{\mu}{\alpha}{\beta}}(\tau)
=\Biglb[R^{{\epsilon}}_{\ {\alpha}{\mu}{\nu}}\xi_{{\epsilon}}
-\delta \Gamma_{{\nu}{\mu}{\alpha}}^R\Bigrb]_{;{\beta}}+
D_{{\nu}{\mu}{\alpha}{\beta}}\,.
\end{equation}
Here all the quantities are evaluated on the worldline, 
$\delta \Gamma_{\nu\mu\alpha}^{R}=\frac{1}{2}(h^{R}_{\mu\nu;\alpha}+h^{R}_{\nu\alpha;\mu}-
h^{R}_{\mu\alpha;\nu})$, and  
$D_{{\nu}{\mu}{\alpha}{\beta}}$ is a tensor field which is defined 
on the worldline (see Appendix C).  
Substituting Eq. (\ref{gradgradgradxi}) 
into Eq.(\ref{gradgradhr}) gives
\begin{equation}\label{hrfd}
 h^{R(F)}_{{\mu}{\nu};{\alpha}{\beta}}
=2D_{({\mu}{\nu}){\alpha}{\beta}}\,.
\end{equation}

Unfortunately, $D_{{\nu}{\mu}{\alpha}{\beta}}$ is not completely arbitrary 
and its values are constrained by various identities [e.g. identity (\ref{commute})].
These constrains prevent us from  
annihilating  $h^{R(F)}_{{\mu}{\nu};{\alpha}{\beta}}$
 by invoking a gauge transformation. 
In Appendix C, we choose $D_{{\nu}{\mu}{\alpha}{\beta}}$ such that
  all the required constraints are satisfied.
We shall refer to this specific choice of gauge as extended Fermi gauge.
In this gauge we have
\begin{equation}\label{hrfrrf}
 h^{R(F)}_{{\mu}{\nu};{\alpha}{\beta}}
=P^{\ {\lambda}\ {\eta}\ {\rho}\ {\sigma}}_{{\mu}\ {\nu}\ {\alpha}
\ {\beta}}
\delta R^{R(F)}_{{\lambda}{\eta}{\rho}{\sigma}}
 \,.
\end{equation}
where 
$P^{\ {\lambda}\ {\eta}\ {\rho}\ {\sigma}}_
{{\mu}\ {\nu}\ {\alpha}\ {\beta}}$ 
is a certain combination of $u^{{\mu}}$ and $\delta_{{\mu}}^{{\nu}}$ 
(see Appendix C); and
 $\mu \delta R^{R(F)}_{{\alpha}{\beta}{\gamma}{\delta}}$
 is the linear term in the following expansion in powers of $\mu$
\begin{equation}\label{rrf}
\biglb(g_{\mu\alpha}+\mu h^{R(F)}_{\mu\alpha} \bigrb)
R^\mu_{\ \beta\gamma\delta}[g_{\mu\nu}+\mu h^{R(F)}_{\mu\nu}]-
g_{\mu\alpha}R^\mu_{\ \beta\gamma\delta}[g_{\mu\nu}]=
\mu \delta R^{R(F)}_{{\alpha}{\beta}{\gamma}{\delta}}+O(\mu^2)\,.
\end{equation}
Here the Riemann tensors are evaluated using the metric inside the squared brackets.
Meaning that  $\mu \delta R^{R(F)}_{{\alpha}{\beta}{\gamma}{\delta}}$
is the linear perturbation of the Riemann tensor, evaluated with 
the metric $g_{\mu\nu}+\mu h^{R(F)}_{\mu\nu}$. 
Notice that  $\delta R^{R(F)}_{{\alpha}{\beta}{\gamma}{\delta}}$
 has the same algebraic
symmetries as the Riemann tensor. Furthermore, in Appendix C we show that 
in vacuum $\delta R^{R(F)}_{\hat{\alpha}\hat{\beta}\hat{\gamma}\hat{\delta}}$
is traceless. We now follow the arguments in Sec. IIA. There we showed 
that in vacuum no vector can be constructed from the tensors
$R_{\hat{\alpha}\hat{\beta}\hat{\gamma}\hat{\delta}}$, $u^{\hat{\mu}}$, $\varepsilon^{\hat{\alpha}\hat{\beta}\hat{\gamma}\hat{\delta}}$, and 
$g_{\hat{\mu}\hat{\nu}}$. Similarly, the above mentioned 
properties  of $\delta R^{R(F)}_{\hat{\alpha}\hat{\beta}\hat{\gamma}\hat{\delta}}$ imply that
 in vacuum no vector can be constructed from  
$\delta R^{R(F)}_{ \hat{\lambda}\hat{\eta}\hat{\rho}\hat{\sigma} }$,
 $u^{\hat{\alpha}}$,  $\varepsilon^{\hat{\alpha}\hat{\beta}\hat{\gamma}\hat{\delta}}$, 
and $g_{ \hat{\alpha}\hat{\beta} }$. Employing 
Eq. (\ref{hrfrrf}) we now find that in the extended Fermi gauge we have 
\begin{equation}\label{f2a0}
f^{\mu}_{(2)A}=0\,.
\end{equation}

\subsection{Second term $f^{\mu}_{(2)B}$} 

Here we focus on the second term
$f^{\mu}_{(2)B}[g_{\alpha\beta},z_G(\tau),\delta l_{\gamma\delta}]$ in Eq. (\ref{f2ab}).
Following Sec. II,  we decompose $\delta l_{\gamma\delta}$, in a local neighborhood of  
$z_G(\tau)$, into a generalized instantaneous piece  $\delta l^{I}_{\mu\nu}$ 
and a sufficiently regular piece  $ \delta l^{SR}_{\mu\nu}$, reading
\begin{equation}\label{dlsisr}
\delta l_{\mu\nu}= \delta l^{I}_{\mu\nu}+ \delta l^{SR}_{\mu\nu}\,.
\end{equation}
Here $\delta l^{I}_{\mu\nu}$  is defined as a piece 
that admits the following a local expansion  on $\Sigma({\hat{\tau}})$
\begin{equation}\label{lidef}
\delta l^{I}_{\alpha\beta}(x)=
\bar{g}^{\ \hat{\alpha}}_{\alpha} \bar{g}^{\ \hat{\beta}}_{\beta}
\Biglb[e^{(0)}_{\hat{\alpha}\hat{\beta}}+
e^{(1)}_{\hat{\alpha}\hat{\beta}} \varepsilon+
e^{(2)}_{\hat{\alpha}\hat{\beta}} \varepsilon\log{\varepsilon}
\Bigrb]\,,
\end{equation}
where the expansion coefficients 
$\{e^{(0)}_{\hat{\alpha}\hat{\beta}},
e^{(1)}_{\hat{\alpha}\hat{\beta}},e^{(2)}_{\hat{\alpha}\hat{\beta}}
\}$
 are composed only from the quantities in the following list:
background tensors (at $\hat{z}$), $u^{\hat{\mu}}$, $\nabla_{\hat{\alpha}}\varepsilon$,
 $\delta R^{R(F)}_{\hat{\lambda}\hat{\eta}\hat{\rho}\hat{\sigma}}$, and numerical
coefficients.
We define a sufficiently regular piece $\delta l^{SR}_{\mu\nu}$ 
such that its first-order covariant
derivative $\nabla_{\alpha}\delta l^{SR}_{\mu\nu}$ is continuous in a local 
neighborhood of $z_G$ and its higher-order derivatives are not too singular  
(More precisely, we demand that 
$\varepsilon^{-1+n}\nabla_{\delta_n}...\nabla_{\delta_1} \delta l^{SR}_{\mu\nu;\gamma}$, where  
$n\ge 1$, remains bounded as $x\rightarrow \hat{z}$.). 

We initially assume that decomposition (\ref{dlsisr}) exists.
Later, in Sec. IIIF, we shall provide a specific prescription for its
 construction. As in  Sec. IID, the properties of  
decomposition  (\ref{dlsisr}) imply that this decomposition in nonunique.
Following Eq. (\ref{anotherdec}), we can generate a family of 
 decompositions by
invoking the transformation $\delta l^{I}_{\mu\nu}\rightarrow \delta l^{I}_{\mu\nu}+j_{\mu\nu}$, 
 $\delta l^{SR}_{\mu\nu}\rightarrow \delta l^{SR}_{\mu\nu}-j_{\mu\nu}$, where 
$j_{\mu\nu}$ satisfies the definitions of both $\delta l^{I}_{\mu\nu}$ 
and  $\delta l^{SR}_{\mu\nu}$.
Following the discussion in  Sec. IID reveals that the non-uniqueness of 
decomposition  (\ref{dlsisr}) does not affect 
our final expression for the second-order self-force [see Eq. (\ref{f2final}) below].

Using Eqs. (\ref{f2ab},\ref{f2a0},\ref{dlsisr}) together with the fact that 
$f^\mu_{(2)}$ must be linear in the second-order perturbations we find that 
we can decompose $f^\mu_{(2)}$ as follows
\begin{equation}\label{f2dec}
f_{(2)}^\mu=f^{\mu}_{(2)B}
[g_{\alpha\beta},z_G(\tau),\delta l_{\gamma\delta}]=f^{\mu}_{(2)I}
[g_{\alpha\beta},z_G(\tau),\delta l^{I}_{\gamma\delta}]+
f^{\mu}_{(2)SR}
[g_{\alpha\beta},z_G(\tau),\delta l^{SR}_{\gamma\delta}]\,.
\end{equation}

\subsubsection {Instantaneous  piece}

First we consider the instantaneous piece $f^{\mu}_{(2)I}
[g_{\alpha\beta},z_G(\tau),\delta l^{I}_{\gamma\delta}]$ in Eq. (\ref{f2dec}). 
The quantities $g_{\alpha\beta},z_G(\tau),\delta l^{I}_{\gamma\delta}$ in this expression 
together with the definition 
of  $\delta l^{I}_{\mu\nu}$ reveal that the tensorial constituents of  
$f^{\mu}_{(2)I}$ are: the background tensors, the four velocity, the Levi-Civita tensor, 
and the tensor $\delta R^{R(F)}_{\hat{\lambda}\hat{\eta}\hat{\rho}\hat{\sigma}}$. 
Recall that in Sec. IIIC we showed that in vacuum these tensors can not 
be combined to give a vector with the desired dimensionality of 
$(Length)^{-3}$. We therefore conclude that 
\begin{equation}\label{f2inst}
f^{\mu}_{(2)I}=0 \,.
\end{equation}

\subsubsection {Sufficiently regular piece}

Next we consider the sufficiently regular piece $f^{\hat{\mu}}_{(2)SR}$.
Following the analysis in Sec. IIB, we list 
the tensorial constituents of 
$f^{\hat{\mu}}_{(2)SR}[g_{\alpha\beta},z_G(\tau),\delta l^{SR}_{\gamma\delta}]$.
The explicit dependence of $f^{\hat{\mu}}_{(2)SR}$ on  the quantities 
$\{g_{\alpha\beta},z_G(\tau)\}$ 
implies that  the background tensors and the four velocity 
must be included in
the our list of the tensorial constituents of $f^{\hat{\mu}}_{(2)SR}$. In addition,  
this list must also include the Levi-Civita tensor.
Recall that in Sec. IIIC we showed  that in vacuum these tensors can not 
be combined to give a non-vanishing vector with a  dimension of
 $(Length)^{-3}$. Therefore, each term in the expression for $f^{\hat{\mu}}_{(2)SR}$ must 
 depend (linearly) on $\delta l^{SR}_{{\mu}{\nu}}$. 

Since   $\delta l^{SR}_{{\mu}{\nu}}$ is sufficiently regular it admits the following
expansion on $\Sigma(\hat{\tau})$
\begin{equation}\label{lsrexp}
 \delta l^{SR}_{\alpha\beta}(x)=
\bar{g}^{\ \hat{\alpha}}_{\alpha} \bar{g}^{\ \hat{\beta}}_{\beta}
\Biglb[  \delta l^{SR}_{\hat{\alpha}\hat{\beta}}-
 \delta l^{SR}_{\hat{\alpha}\hat{\beta};\hat{\gamma}}
\varepsilon\varepsilon^{;\hat{\gamma}}+O(\varepsilon^2)\Bigrb]\,.
\end{equation}
Recall that  $ \delta l^{SR}_{{\alpha}{\beta}}$
has  a dimension of $(Length)^{-2}$. Thus  $\delta l^{SR}_{\hat{\alpha}\hat{\beta};\hat{\gamma}\hat{\delta}_1...\hat{\delta}_n}$ 
$n\ge 1$  must have dimensions of  $(Length)^{-3-n}$. 
By dimensionality, these higher-order derivatives  must be 
excluded from our list \cite{nonreg}.
Consequently,  each term in the expression of  $f^{\hat{\mu}}_{(2)SR}$   
must  depend  on   $\delta l^{SR}_{\hat{\alpha}\hat{\beta}}$ or 
 $\delta l^{SR}_{\hat{\alpha}\hat{\beta};\hat{\gamma}}$. 
To eliminate all the possibilities of constructing a vector with a dimension
of  $(Length)^{-3}$ we employ a purely second-order gauge transformation 
 of the form $x^\mu\rightarrow x^\mu - \mu^2 \xi_{(2)}^\mu$. 
The second-order perturbations in this second-order Fermi  
gauge are given by $l^F_{\mu\nu}=l_{\mu\nu}+\xi_{(2)\mu;\nu}+\xi_{(2)\nu;\mu}$.
Similar to Sec. IIB, we include the entire gauge transformation
in the definition of the new sufficiently regular piece 
$\delta l^{SR(F)}_{\mu\nu}\equiv \delta l^{SR}_{\mu\nu}+
\xi_{(2)\mu;\nu}+\xi_{(2)\nu;\mu}$ and demand that  $\xi_{(2)\mu;\nu\rho}$ will be  continuous
 in a local neighborhood of the worldline and thereby ensure that 
$\delta l^{SR(F)}_{\mu\nu}$
will satisfy the conditions of a sufficiently regular piece.
In the second-order Fermi gauge Eq.(\ref{f2dec})  reads
\begin{equation}\label{f2decfermi}
f_{F(2)}^\mu=
f^{\mu}_{(2)I}
[g_{\alpha\beta},z_G(\tau),\delta l^{I}_{\gamma\delta}]+
f^{\mu}_{(2)SR(F)}
[g_{\alpha\beta},z_G(\tau),\delta l^{SR(F)}_{\gamma\delta}]\,.
\end{equation}
where $f^\mu_{F(2)}$ is the second-order self-force in the second-order Fermi gauge, 
and $f^\mu_{(2)SR(F)}$ is its corresponding sufficiently regular piece.
Following Sec. IIB we impose the following gauge conditions along the world line $z_G(\tau)$
\begin{equation}\label{fgauge2}
\left[\delta l^{SR(F)}_{\mu\nu}\right]_{z_G(\tau)}=0\,,\,
\left[\delta l^{SR(F)}_{\mu\nu;\rho} \right]_{z_G(\tau)}=0\,.
\end{equation}
Recall that each term in  $f^{\hat{\mu}}_{(2)SR}$ must depend on 
  $\delta l^{SR}_{\hat{\alpha}\hat{\beta}}$ or 
 $\delta l^{SR}_{\hat{\alpha}\hat{\beta};\hat{\gamma}}$. 
In complete analogy, each term in $f^{\hat{\mu}}_{(2)SR(F)}$ must depend      
 on $\delta l^{SR(F)}_{\hat{\alpha}\hat{\beta}}$ 
or $\delta l^{SR(F)}_{\hat{\alpha}\hat{\beta};\hat{\gamma}}$. Notice that these  coefficients vanish 
by virtue of Eq. (\ref{fgauge2}). We therefore conclude  that    
\begin{equation}\label{f2sr}
f^{{\mu}}_{(2)SR(F)}=0\,.
\end{equation}

\subsection {Second-order self-force}

Combining Eqs. (\ref{f2inst},\ref{f2sr}) together with Eq. (\ref{f2decfermi}) 
we find that in the second-order Fermi gauge the second-order self-force vanishes, namely 
\begin{equation}\label{f2fermi}
f^{\mu}_{F(2)}=0\,.
\end{equation}
We now invoke an inverse gauge
transformation of the form  $x^\mu\rightarrow x^\mu + \mu^2 \xi_{(2)}^\mu$
and calculate the second-order self-force in our original 
second-order gauge. For this purpose we need to generalize 
Barack and Ori  gauge transformation formula of Ref. \cite{BOGAUGE}
[see Eqs. (\ref{ftrans},\ref{deltaf1})] to second order.
The generalization turns out to be trivial. The first-order  expression (\ref{deltaf1})
is obtained by calculating the leading order
acceleration of the worldline which is induced by a first-order gauge transformation. 
Here we consider a purely second-order gauge transformation of the form
  $x^\mu\rightarrow x^\mu - \mu^2 \xi_{(2)}^\mu$,
but similar to Ref. \cite{BOGAUGE}, we are still calculating the leading order acceleration
 which this transformation induces. Therefore, the  final expression has    
the same form as  Eqs. (\ref{ftrans},\ref{deltaf1}), with the following substitutions
$\xi^\mu \rightarrow\xi_{(2)}^\mu$, 
$f^{\alpha}_{(1)}\rightarrow f^{\alpha}_{(2)}$, 
$\delta f^{\alpha}_{(1)}\rightarrow \delta f^{\alpha}_{(2)}$.

The final piece of the calculation directly follows from the analysis 
in Sec. IID. We use the 
gauge conditions (\ref{fgauge2}) to derive an expression for $\ddot{\xi}_{(2)}^\mu$, and use 
this expression to find $\delta f^{\alpha}_{(2)}$ and $f^{\mu}_{(2)}$.
This calculation produces our main result which is the following
 expression for the 
second order self-force, reading
\begin{equation}\label{f2final}
\mu^3 f^{\mu}_{(2)}=\mu^3 K^{\mu\alpha\beta\gamma} \delta l^{SR}_{\alpha\beta;\gamma}\,.
\end{equation}
Here the first-order metric perturbations satisfy the conditions 
of the extended Fermi gauge, the second-order 
metric perturbations are constructed using the prescription of 
Sec. IIIA, and the piece $\delta\bar{l}_{\mu\nu}$ is decomposed 
according to Eq.\, (\ref{dlsisr}).
To complete the derivation we now provide a prescription for the construction 
of this decomposition.

\subsection {Specific decomposition of $\delta l_{\mu\nu}$}

To construct decomposition (\ref{dlsisr}) 
we need to study the singular behavior of 
$\delta l_{\mu\nu}$ in the vicinity of the worldline. For this purpose,  
we first study the singular behavior of the source term 
 $-2\delta S_{\alpha\beta}$ of Eq. (\ref{deltah2final}). 
From its definition $\delta S_{\alpha\beta}$ has the schematic form 
$\delta S=\nabla h^F\nabla h^F\, \&\, h^F\nabla\nabla h^F$, where $h^F$ schematically 
denotes the first-order metric-perturbations in the extended Fermi-gauge, and $''\&''$ denotes
``and terms of the form ...''. Recall that $h^F$
diverges as  $\varepsilon^{-1}$ as $\varepsilon\rightarrow 0 $. 
Therefore, the schematic form of $\delta S_{\alpha\beta}$ 
suggests that $\delta S_{\alpha\beta}$
should diverge as $\varepsilon^{-4}$. However, 
  Refs. \cite{eran1,eran2} show that the leading singular terms in   $\delta S_{\alpha\beta}$
cancel out and $\delta S_{\alpha\beta}$ diverges only as  $\varepsilon^{-2}$.
Formally expanding  $-2\delta S_{\alpha\beta}$ on $\Sigma(\hat{\tau})$ gives
\begin{equation}\label{dsexpand}
-2\delta S_{\alpha\beta}=\bar{g}_{\alpha}^{\ \hat{\alpha}} \bar{g}_{\beta}^{\ \hat{\beta}}
[\varepsilon^{-2}A_{\hat{\alpha}\hat{\beta}}+\varepsilon^{-1}B_{\hat{\alpha}\hat{\beta}}
+O(\varepsilon^0)]\,,
\end{equation}
where the coefficients $A_{\hat{\alpha}\hat{\beta}},B_{\hat{\alpha}\hat{\beta}}$ are 
independent of $\varepsilon$. Notice that this equation  uniquely defines 
$A_{\hat{\alpha}\hat{\beta}}$ and $B_{\hat{\alpha}\hat{\beta}}$.
We now list the tensorial constituents  
of  the coefficients $A_{\hat{\alpha}\hat{\beta}},B_{\hat{\alpha}\hat{\beta}}$. 
To construct this list, we substitute 
decomposition (\ref{dwfermi}) into the  schematic expression for 
$\delta S_{\alpha\beta}$ and use 
Eq. (\ref{hrfexp}), Eq. (\ref{hsexpand}), and Eq. (\ref{generalform}) 
in Appendix B. We find that 
  each term in the expression for $A_{\hat{\alpha}\hat{\beta}}$
  must be linear in $R_{\hat{\lambda}\hat{\eta}\hat{\rho}\hat{\sigma}}$, and each term 
in the expression 
for $B_{\hat{\alpha}\hat{\beta}}$ must be either linear in
 $\delta R^{R(F)}_{\hat{\lambda}\hat{\eta}\hat{\rho}\hat{\sigma}}$ or linear in 
 $\nabla_{\hat{\alpha}}R_{\hat{\lambda}\hat{\eta}\hat{\rho}\hat{\sigma}}$. 
In addition to these tensors, the coefficients 
$A_{\hat{\alpha}\hat{\beta}}$ and $B_{\hat{\alpha}\hat{\beta}}$ 
 include only combinations of the following dimensionless quantities:
$u^{\hat{\mu}}$, $g_{\hat{\mu}\hat{\nu}}$, $\nabla_{\hat{\alpha}}\varepsilon$, 
and numerical coefficients.

To construct $\delta \bar{l}^{I}_{\alpha\beta}$ we first 
construct a solution to the following equation, defined in a local
neighborhood of the worldline
\begin{equation}\label{deltalieq}
\Box \bar{\varphi}_{\alpha\beta}+2R^{\eta\ \rho}_{\ \alpha\ \beta}
\bar{\varphi}_{\eta\rho}= 
\bar{g}_{\alpha}^{\ \hat{\alpha}} \bar{g}_{\beta}^{\ \hat{\beta}}
[\varepsilon^{-2}A_{\hat{\alpha}\hat{\beta}}+\varepsilon^{-1}B_{\hat{\alpha}\hat{\beta}}
+O_a(\varepsilon^0)]\,.
\end{equation}
Here $O_a(\varepsilon^0)$ denotes 
an arbitrary quantity of order $\varepsilon^0$, meaning that we do not 
restrict the values of the $O(\varepsilon^0)$ terms on the right hand side. 
To construct a solution in the
desired form we substitute
\begin{equation}\label{phihathat}
\bar{\varphi}_{\alpha\beta}=\bar{g}_{\alpha}^{\ \hat{\alpha}} \bar{g}_{\beta}^{\ \hat{\beta}}
\bar{\varphi}_{\hat{\alpha}\hat{\beta}}
\end{equation}
into Eq. (\ref{deltalieq}), and solve directly for $\bar{\varphi}_{\hat{\alpha}\hat{\beta}}$.
Notice that 
$A_{\hat{\alpha}\hat{\beta}},B_{\hat{\alpha}\hat{\beta}}$, and $\bar{\varphi}_{\hat{\alpha}\hat{\beta}}$ transform as scalars under a coordinate transformation at $x$. 
It is useful to decompose  
these quantities into scalar spherical harmonics
\begin{eqnarray}\label{phidec}
&&\bar{\varphi}_{\hat{\alpha}\hat{\beta}}=\sum_{l=0}^{\infty}
\sum_{m=-l}^l Y_{lm}(\theta,\varphi)\bar{\phi}_{\hat{\alpha}\hat{\beta}}^{lm}(\varepsilon,t)\,,\\\nonumber
&&A_{\hat{\alpha}\hat{\beta}}=\sum_{l=0}^{\infty}
\sum_{m=-l}^l Y_{lm}(\theta,\varphi)a_{\hat{\alpha}\hat{\beta}}^{lm}(t)\,,\\\nonumber
&&B_{\hat{\alpha}\hat{\beta}}=\sum_{l=0}^{\infty}
\sum_{m=-l}^l Y_{lm}(\theta,\varphi)b_{\hat{\alpha}\hat{\beta}}^{lm}(t)\,.
\end{eqnarray}
To define the variables $(t,\theta,\varphi)$, consider 
 Fermi-normal coordinates based on the worldline, where $x^{a}$ $(a=1,2,3)$ denote  
the spatial coordinates and $t$ denotes the time coordinate. The angular variables are 
defined through the canonical angular parametrization, namely 
$x^a=\varepsilon(\sin\theta\cos\varphi,\sin\theta\sin\varphi,\cos\theta)$.
In Appendix D we construct the following  solution to Eq. (\ref{deltalieq}), reading 
\begin{eqnarray}\label{solforphi}
&&\bar{\phi}_{\hat{\alpha}\hat{\beta}}^{00}=
\frac{1}{2} b_{\hat{\alpha}\hat{\beta}}^{00}\varepsilon\ ,\ l=0\\\nonumber 
&&\bar{\phi}_{\hat{\alpha}\hat{\beta}}^{1m}=
-\frac{1}{2}a_{\hat{\alpha}\hat{\beta}}^{1m}+
\frac{1}{3}b_{\hat{\alpha}\hat{\beta}}^{1m}\varepsilon\log\varepsilon\ , \ l=1
\\\nonumber
&&\bar{\phi}_{\hat{\alpha}\hat{\beta}}^{lm}=
-[l(l+1)]^{-1}a_{\hat{\alpha}\hat{\beta}}^{lm}+
[2-l(l+1)]^{-1}b_{\hat{\alpha}\hat{\beta}}^{lm}\varepsilon\ ,\ l>1\,.
\end{eqnarray}
Notice that the angular dependence of  $A_{\hat{\alpha}\hat{\beta}}$
and $B_{\hat{\alpha}\hat{\beta}}$ is completely described by combinations of 
the quantity $\nabla_{\hat{\alpha}}\varepsilon$. Therefore,  when we calculate
the spherical-harmonics coefficients
$a_{\hat{\alpha}\hat{\beta}}^{lm}$ and $b_{\hat{\alpha}\hat{\beta}}^{lm}$, 
we end up integrating only over combinations of  $\nabla_{\hat{\alpha}}\varepsilon$.
Therefore,  $\bar{\varphi}_{\hat{\alpha}\hat{\beta}}$
is composed form  the same well defined tensorial constituents
 that appear in the expressions of  $A_{\hat{\alpha}\hat{\beta}}$
and $B_{\hat{\alpha}\hat{\beta}}$. In addition,   
$\bar{\varphi}_{\hat{\alpha}\hat{\beta}}$  
also has a non-trivial angular dependence which can be expressed  
using combinations of $\nabla_{\hat{\alpha}}\varepsilon$.
Comparing Eqs. (\ref{phihathat},\ref{phidec},\ref{solforphi}) 
with Eq. (\ref{lidef}) we find that $\bar{\varphi}_{\alpha\beta}$ satisfies the conditions of 
a generalized instantaneous piece. We therefore identify $\bar{\varphi}_{\alpha\beta}$
 with the desired generalized instantaneous piece, namely
\begin{equation}\label{liphi}
\delta \bar{l}^{I}_{\alpha\beta}=\bar{\varphi}_{\alpha\beta}\,.
\end{equation}
By subtracting Eq. (\ref{deltalieq}) (for the above solution $\bar{\varphi}_{\alpha\beta}$) 
from Eq. (\ref{deltah2final})   we obtain
\begin{equation}\label{deltalsreq}
\Box (\delta\bar{l}_{\alpha\beta}-\bar{\varphi}_{\alpha\beta})+2R^{\eta\ \rho}_{\ \alpha\ \beta}
(\delta \bar{l}_{\eta\rho}-\bar{\varphi}_{\eta\rho})= 
\bar{g}_{\alpha}^{\ \hat{\alpha}} \bar{g}_{\beta}^{\ \hat{\beta}}
[O(\varepsilon^0)]\,.
\end{equation}
In Appendix D we use the fact that the source term of Eq. (\ref{deltalsreq}) 
is non-divergent to show that  $\delta\bar{l}_{\alpha\beta}-\bar{\varphi}_{\alpha\beta}$
 satisfies the conditions of a sufficiently regular piece. 
We therefore make the following identification
\[
\delta \bar{l}^{SR}_{\alpha\beta}=
\delta\bar{l}_{\alpha\beta}-\bar{\varphi}_{\alpha\beta}\,,
\]
which completes the construction of decomposition (\ref{dlsisr}).

\subsection{Summary}

Eq. (\ref{f2final}) provides a formula for calculating the 
second-order gravitational self-force in a vacuum background spacetime, given 
the sufficiently regular piece $\delta {l}^{SR}_{\alpha\beta}$.
To use this formula one has to be able 
to calculate both first-order and second-order metric perturbations,  
in specified first and second order gauges. 
We now briefly summarize a prescription 
for the construction of  $\delta {l}^{SR}_{\alpha\beta}$.

The first step in this construction is to  calculate the 
retarded solution of Eq. (\ref{waveeq}) which provides us with $\bar{h}_{\mu\nu}$ --
the (traced reversed) first-order metric-perturbations 
in the Lorenz-gauge. 
The next step is to calculate the gauge vector $\xi_{\mu}$ that generates the 
gauge-transformation 
$h_{\mu\nu}\rightarrow h_{\mu\nu}+\xi_{\nu;\mu}+\xi_{\mu;\nu}$ from  the Lorenz-gauge to 
 the extended Fermi-gauge.  
To be able to calculate $\xi_{\mu}$,  one first has to 
 calculate the regular piece $h^R_{\mu\nu}(z_G)$ together 
with its first and second order derivatives at the worldline.  
This preliminary calculation follows from  Eqs. (\ref{dw},\ref{hs},\ref{gs}).
Next, one follows  the prescription in Sec. IIB [immediately after Eq. (\ref{f1sr})] 
and calculates  $\xi^\mu(z_G)$ together with   
its first and second order derivatives.
 To construct the third order derivative of  $\xi^\mu$  
one should use Eq. (\ref{gradgradgradxi}) together with 
Eqs. (\ref{dfermi1},\ref{dfermi2}) in Appendix C. 
Once $\xi^\mu(z_G)$ together with its first, second 
and third order derivatives on the worldline are obtained, one can use 
these quantities to  
construct a local expansion for  $\xi^\mu$ (this expansion can be 
continued  in an arbitrary manner away from the worldline).

The next step in our construction is to calculate the (traced reversed) 
second-order metric perturbations $\bar{l}_{\mu\nu}$. For this purpose  
one can use the decomposition $\bar{l}_{\mu\nu}=\bar{\psi}_{\mu\nu}+\delta\bar{l}_{\mu\nu}$.
The first piece of this decomposition -- $\bar{\psi}_{\mu\nu}$
 is given by Eq. (\ref{psiexplicit}),  and the second piece -- $\delta\bar{l}_{\mu\nu}$
is equal to the retarded solution of  Eq. (\ref{deltah2final}). 
The piece $\delta\bar{l}_{\mu\nu}$ is further decomposed according to 
 $\delta \bar{l}_{\mu\nu}= \delta \bar{l}^{I}_{\mu\nu}+ \delta \bar{l}^{SR}_{\mu\nu}$.
Eqs. (\ref{dsexpand},\ref{phihathat},\ref{phidec},\ref{solforphi},\ref{liphi}) provides a prescription for 
 the calculation of the desired sufficiently regular 
piece $\delta {l}^{SR}_{\alpha\beta}$ that should be substituted in Eq. (\ref{f2final}).

\acknowledgments

I am grateful to Amos Ori and to Eric Poisson 
for valuable discussions.
This work was supported  by the Natural Sciences and Engineering
Research Council of Canada.

\appendix 

\section{Local expansion of $h^{SR}_{\alpha\beta}$}

In this Appendix we derive Eq. (\ref{hsrexp}), which provides a local 
expansion of a sufficiently-regular tensor-field on a spacelike hypersurface $\Sigma({\hat{\tau}})$. 
Most of our derivation follows a similar derivation in Ref. \cite{PR}.  
We begin our analysis by studying  some of the properties of
the quantities $\varepsilon\varepsilon^{;\hat{\mu}}$ 
and $\bar{g}^{\ \hat{\alpha}}_{\alpha} $ that appear in  Eq. (\ref{hsrexp}).

First we derive an expression for  $\varepsilon\varepsilon^{;\hat{\mu}}$.  
Consider two points $x'$ and $x$, 
and suppose that $x'$ lies in a local neighborhood of $x$, such that within this 
local neighborhood there is a unique geodesic $y(\lambda)$ that connects $x'$ with $x$.
Here  $\lambda$ denotes an affine parameter ranging from $\lambda_0$ to $\lambda_1$, 
where  $x'=y(\lambda_0)$ and  $x=y(\lambda_1)$.  
We denote the square of the invariant length of  $y(\lambda)$ with $2\sigma(x|x')$,  and 
express  $\sigma(x|x')$ as
\begin{equation}\label{sigmadef}
\sigma(x|x')=\frac{1}{2}(\lambda_1-\lambda_0)\int_{\lambda_0}^{\lambda_1}
g_{\mu\nu}(y)\dot{y}^\mu \dot{y}^\nu d\lambda\,,
\end{equation}
where $\dot{y}^\mu=\frac{dy}{d\lambda}$. 
We now calculate $\sigma_{,\mu'}=\frac{\partial \sigma}{\partial x^{\mu'}}$. 
For this purpose we consider an infinitesimal displacement of 
the point $x'$ to a point $x'+\delta x'$, and let $y(\lambda)+\delta y(\lambda)$ denote the unique 
 geodesic connecting $x'+\delta x'$ with $x$. On this geodesic we scale the affine parameter $\lambda$ to
run from $\lambda_0$ to $\lambda_1$, such that  
 $x'+\delta x'=y(\lambda_0)+\delta y(\lambda_0)$ and   $\delta y(\lambda_1)=0$.
Expanding   $\delta \sigma=\sigma(x|x'+\delta x')-\sigma(x|x')$ to the first order in the variation, 
and using integration by parts we obtain 
\[
\delta \sigma=(\lambda_1-\lambda_0)\biglb[ g_{\mu\nu}(y) \dot{y}^\nu \delta y^\mu \bigrb]^{\lambda_1}_{\lambda_0} - 
(\lambda_1-\lambda_0)\int_{\lambda_0}^{\lambda_1} \frac{D\dot{y}_{\mu}}{D\lambda} d\lambda \delta y^\mu+O(\delta y^2)\,.
\]
Recalling that  $y(\lambda)$ is a geodesic and noting that  $\delta y(\lambda_1)=0$ we obtain
\begin{equation}\label{gradsigma}
\sigma_{,\mu'}=-(\lambda_1-\lambda_0)g_{\mu'\nu'}\dot{y}^{\nu'}\,.
\end{equation}
Hereafter we shall specialize to the space-like  geodesic that connects the points $\hat{z}$ and $x$ on 
$\Sigma(\hat{\tau})$. Here we have 
$2\sigma(y|\hat{z})=\varepsilon^2(y|\hat{z})$. Employing  Eq. (\ref{gradsigma}) and using 
$\varepsilon(y)$ as the affine parameter of the geodesic, we find that 
\begin{equation}\label{epsgradeps}
\varepsilon \nabla ^{\hat{\mu}} \varepsilon= -\varepsilon \frac{dy^{\hat{\mu}}} {d \varepsilon}\,,
\end{equation}
which implies that $ \nabla ^{\hat{\mu}} \varepsilon$ is a unit vector, tangent to  the geodesic $y(\varepsilon)$.

Next we study the properties of the parallel propagator $\bar{g}^{\ \hat{\alpha}}_{\alpha} $. From its definition 
 $\bar{g}^{\ \hat{\alpha}}_{\alpha} $  transports an arbitrary vector $A_{\hat{\alpha}}$ to a vector 
\begin{equation}\label{pp}
 A_\mu(y)=\bar{g}^{\ \hat{\alpha}}_{\mu}(y|\hat{z})A_{\hat{\alpha}}(\hat{z})\,,
\end{equation}
by a parallel propagation of the vector  $A_{\hat{\alpha}}$ on the geodesic $y(\varepsilon)$, 
implying  that
\begin{equation}\label{pplimit}
\lim_{\varepsilon\rightarrow 0} 
\bar{g}^{\ \hat{\alpha}}_{\mu}[y(\varepsilon)|\hat{z}]=\delta^{\ \hat{\alpha}}_{\hat{\mu}}\,.
\end{equation}
Since  $ \frac{D}{D\varepsilon}A_{\mu}=0 $ we find from Eq. (\ref{pp}) that 
\begin{equation}\label{gradpp}
(\bar{g}^{\ \hat{\alpha}}_{\mu})_{;\rho}\frac{dy^\rho}{d\varepsilon}=0\,.
\end{equation}
The geometric content of  $\bar{g}^{\ \hat{\alpha}}_{\alpha} $ implies 
that  $\bar{g}^{\ \hat{\alpha}}_{\alpha} \bar{g}^{{\beta}}_{\ \hat{\alpha}}=\delta ^{{\beta}} _{\alpha} $,   and 
$\bar{g}^{\ \hat{\alpha}}_{\alpha} \bar{g}^{{\alpha}}_{\ \hat{\beta}}=\delta ^{\hat{\alpha}} _{\hat{\beta}} $. 

We now  construct  a local-expansion for 
 $h^{SR}_{\alpha\beta}(x)$ on $\Sigma({\hat{\tau}})$, this 
construction follows directly from the expansion of 
  $h^{SR}_{\alpha\beta}(y)$ on the geodesics $y(\varepsilon)$.
First we introduce the following 
 bi-tensor field on $y(\varepsilon)$ 
\begin{equation}\label{bdef}
B_{\hat{\alpha}\hat{\beta}}[\hat{z},y]=\bar{g}_{\ \hat{\alpha}}^{\alpha}(y|\hat{z})
 \bar{g}_{\ \hat{\beta}}^{\beta} (y|\hat{z})h^{SR}_{\alpha\beta}(y)\,.
\end{equation}
Notice that  $B_{\hat{\alpha}\hat{\beta}}$ 
transforms as a scalar under a coordinate transformation at $y$. 
Expanding  $B_{\hat{\alpha}\hat{\beta}}(\varepsilon)=B_{\hat{\alpha}\hat{\beta}}[\hat{z},y(\varepsilon)]$ 
in a Taylor series on the geodesic $y(\varepsilon)$, with $\hat{z}$ fixed,  gives 
\begin{equation}\label{btaylor}
B_{\hat{\alpha}\hat{\beta}}(\varepsilon)=
B_{\hat{\alpha}\hat{\beta}}(0)+
\varepsilon\left[\frac { dB_{\hat{\alpha}\hat{\beta}}} {d\varepsilon}\right]_{\varepsilon=0}
+O(\varepsilon^2)\,.
\end{equation}
Eqs. (\ref{pplimit},\ref{bdef}) imply that 
\begin{equation}\label{b0}
B_{\hat{\alpha}\hat{\beta}}(0)=h^{SR}_{\hat{\alpha}\hat{\beta}}(\hat{z})\,.
\end{equation}
Using Eqs.(\ref{gradpp},\ref{bdef}) we find that 
\[
\frac {d B_{\hat{\alpha}\hat{\beta}}} {d\varepsilon}=
B_{\hat{\alpha}\hat{\beta};\gamma}\frac{dy^\gamma}{d\varepsilon}=
\bar{g}_{\ \hat{\alpha}}^{\alpha}(y|\hat{z})
 \bar{g}_{\ \hat{\beta}}^{\beta} (y|\hat{z})h^{SR}_{\alpha\beta;\gamma}(y)\frac{dy^\gamma}{d\varepsilon}
\]
 Employing Eqs.(\ref{epsgradeps},\ref{pplimit}) we obtain 
\begin{equation}\label{gradb0}
\left[\frac { dB_{\hat{\alpha}\hat{\beta}}} {d\varepsilon}\right]_{\varepsilon=0}=
-h^{SR}_{\hat{\alpha}\hat{\beta};\hat{\gamma}}(\hat{z})\varepsilon^{;\hat{\gamma}}\,.
\end{equation}
We can now rewrite $B_{\hat{\alpha}\hat{\beta}}[\varepsilon(y)]$ for $y=x$ 
by substituting Eqs. (\ref{b0},\ref{gradb0}) into Eq. (\ref{btaylor}). Using  this 
expression for   $B_{\hat{\alpha}\hat{\beta}}$ together with 
Eq. (\ref{bdef}) we finally obtain Eq. (\ref{hsrexp}), reading 
\[
h^{SR}_{\alpha\beta}(x)=
\bar{g}^{\ \hat{\alpha}}_{\alpha} (x|\hat{z})\bar{g}^{\ \hat{\beta}}_{\beta}(x|\hat{z})
\Biglb[ h^{SR}_{\hat{\alpha}\hat{\beta}}-
h^{SR}_{\hat{\alpha}\hat{\beta};\hat{\gamma}}
\varepsilon\varepsilon^{;\hat{\gamma}}+O(\varepsilon^2)\Bigrb]\,,
\]
This expansion takes a very simple form in Fermi-normal coordinates based on $z_G$ 
(For the definition and properties of these coordinates see e.g. \cite{PR}.).
In these coordinates we have 
$\bar{g}^{\ \hat{\alpha}}_{\alpha} =\delta^{\ \hat{\alpha}}_{\alpha}+O(\varepsilon^2) $, 
and Eq. (\ref{hsrexp}) is reduced to
\begin{equation}\label{fnchsr}
h^{SR}_{\alpha\beta}(x)\stackrel{*}{=} h^{SR}_{\hat{\alpha}\hat{\beta}}(t)+
h^{SR}_{\hat{\alpha}\hat{\beta},a}(x^a=0,t) x^a+O(\varepsilon^2)\,.
\end{equation}
Here $\stackrel{*}{=}$ denotes equality in Fermi normal coordinates, $t$ denotes 
the time coordinate, $x^a$ denote the spatial coordinates, where 
Latin indices run from $1$ to $3$. Eq. (\ref{fnchsr})  clearly shows 
that  $h^{SR}_{\alpha\beta}$ is a regular tensor-field in a local-neighborhood of the world-line.  
In Fermi normal coordinates it also easy to evaluate the first-order covariant 
derivatives of Eq. (\ref{fnchsr}) since 
$\biglb[\nabla_{\alpha}\bigrb]_{\hat{z}}\stackrel{*}{=}\biglb[\partial_{\alpha}\bigrb]_{\hat{z}}$.

\section {Local expansion of  $\bar{h}^S_{\mu\nu}$ }

In this Appendix we study the general form 
of a local expansion of $\bar{h}^S_{\mu\nu}$ (For
 more details on the construction method of 
such local expansions see e.g. Ref. \cite{PR}.).
First, we substitute Eq. (\ref{gs})
into Eq. (\ref{hs}) and obtain
\begin{equation}\label{hs1}
\bar{h}^S_{\mu\nu}(x)=2U_{\mu\nu}[x|z_G(\tau^{-})]
\left(\dot{\sigma}_{\tau^-}\right)^{-1}-
2U_{\mu\nu}[x|z_G(\tau^{+})]\left(\dot{\sigma}_{\tau^+}\right)^{-1}
+2\int_{\tau^{-}}^{\tau^{+}}V_{\mu\nu}[x|z_G(\tau)]d\tau\,.
\end{equation}
Here we introduced the following notation: $\sigma_\tau=\sigma(z_G(\tau)|x)$, 
$\dot{\sigma}_{\tau^\pm}={(\frac{d\sigma}{d\tau})}_{\tau^\pm}$,
$U_{\mu\nu}[x|z_G(\tau)]=
U_{\mu\nu\alpha\beta}[x|z_G(\tau)]u^{\alpha}(\tau)u^{\beta}(\tau)$,   
 $V_{\mu\nu}[x|z_G(\tau)]=
V_{\mu\nu\alpha\beta}[x|z_G(\tau)]u^{\alpha}(\tau)u^{\beta}(\tau)$; 
 the retarded and advanced times are denoted $\tau^-$ and $\tau^+$, respectively.
These times satisfy $\sigma[x|z_G(\tau^{\mp})]=0$, where $\tau^+>\hat{\tau}$ 
and  $\tau^-<\hat{\tau}$.
Introducing the notation  $\Delta {\tau^{\mp}}=|\tau^{\mp}-\hat{\tau}|$,  
we expand the quantities 
in Eq. (\ref{hs1}) in  Taylor series around $\hat{\tau}$ and obtain
\begin{eqnarray}\label{timeexp}
&&\dot{\sigma}_{\tau^\mp}=
\dot{\sigma}_{\hat{\tau}}\mp \ddot{\sigma}_{\hat{\tau}} \Delta \tau^{\mp}
+ (1/2)\dddot{\sigma}_{\hat{\tau}} (\Delta \tau^{\mp})^2
 +O [(\Delta \tau^{\mp})^3]\,,\\\nonumber
&&U_{\mu\nu}[x|z_G(\tau^{\mp})]=
U_{\mu\nu}[x|\hat{z}]\mp\Delta \tau^{\mp}\dot{U}_{\mu\nu}[x|\hat{z}]
+(1/2){(\Delta \tau^{\mp})}^2\ddot{U}_{\mu\nu}[x|\hat{z}]+O[(\Delta {\tau^{\mp})}^3]\,,\\\nonumber
&&V_{\mu\nu}[x|z_G(\tau)]=
V_{\mu\nu}[x|\hat{z}]+(\tau-\hat{\tau})\dot{V}_{\mu\nu}[x|\hat{z}]
+(1/2)(\tau-\hat{\tau})^2 \ddot{V}_{\mu\nu}[x|\hat{z}]+O[(\tau-\hat{\tau})^3]\,.
\end{eqnarray}
From the relation  $\sigma[x|z_G(\tau^{\mp})]=0$ we find that  
\begin{equation}\label{tineps}
\Delta {\tau^{\mp}}=\varepsilon(1-
\frac{1}{6}\varepsilon^2 
R_{\hat{\alpha}\hat{\gamma}\hat{\beta}\hat{\delta}}\varepsilon^{;\hat{\gamma}}
\varepsilon^{;\hat{\delta}} u^{\hat{\alpha}} u^{\hat{\beta}})+O(\varepsilon^4)\,.
\end{equation}
Notice that expansion (\ref{tineps}) was derived using only geometrical consideration. 
Therefore, the higher order terms in this expansion must also have a geometric content. 
Substituting Eq. (\ref{timeexp}) together with Eq. (\ref{tineps}) into Eq. (\ref{hs1})
one obtains an expansion in powers of $\varepsilon$  
involving only quantities defined on $\Sigma (\hat{\tau})$. Each of these 
quantities can be further expanded in powers of  $\varepsilon$, 
for example  
\[
U_{\alpha\beta}[x|\hat{z}]=
\bar{g}^{\ \hat{\alpha}}_{\alpha} \bar{g}^{\ \hat{\beta}}_{\beta}
[u_{\hat{\alpha}\hat{\beta}}-
\varepsilon \varepsilon^{;\hat{\gamma}}u_{\hat{\alpha}\hat{\beta};\hat{\gamma}}+
\frac{1}{2}\varepsilon^2 \varepsilon^{;\hat{\gamma}} \varepsilon^{;\hat{\delta}}
u_{\hat{\alpha}\hat{\beta};(\hat{\gamma}\hat{\delta})}
 +O(\varepsilon^3)]\,.
\]
In general the coefficients in these expansions  are obtained from
  coincidence limits of the expanded quantity and its covariant 
derivatives .
In this example  the tensors
$u_{\hat{\alpha}\hat{\beta}}$,   
$u_{\hat{\alpha}\hat{\beta};\hat{\gamma}}$, and 
$u_{\hat{\alpha}\hat{\beta};(\hat{\gamma}\hat{\delta})}$ are equal to 
 the coincidence limits (as $x\rightarrow \hat{z}$) of  $U_{{\alpha}{\beta}}$,   
$U_{{\alpha}{\beta};{\gamma}}$, and $U_{{\alpha}{\beta};({\gamma}{\delta})}$, 
 respectively.    
These
  coincidence limits  have 
a geometrical content and they 
are obtained from the definitions of  $\sigma$, $U_{\mu\nu\alpha\beta}$ 
 and  $V_{\mu\nu\alpha\beta}$,  see e.g. \cite{PR}.
The form of the above mention expansions reveals that the expansion of 
$\bar{h}^S_{\mu\nu}(x)$ has the following form 
\begin{equation}\label{generalform}
\bar{h}^S_{\mu\nu}(x)=
\bar{g}^{\ \hat{\alpha}}_{\alpha} \bar{g}^{\ \hat{\beta}}_{\beta}
\Sigma_{n=-1}^N
f^{(n)}_{\hat{\alpha}\hat{\beta}} \varepsilon^{n}
\,.
\end{equation}
Here the expansion coefficients $f^{(n)}_{\hat{\alpha}\hat{\beta}}$ 
are composed only from the following geometrical quantities: 
background tensors, $u^{\hat{\alpha}}$, $\varepsilon^{;\hat{\alpha}}$; 
together with  numerical coefficients.
The first two terms in expansion (\ref{generalform})
 are provided by Eq. (\ref{hsexpand}).

\section{Extended Fermi gauge}

In this Appendix,  we provide a prescription for the construction of the extended
Fermi gauge. 
In addition, we derive Eq. (\ref{hrfrrf})  which provides an expression for 
$h^{R(F)}_{{\mu}{\nu};{\alpha}{\beta}}$
 in this gauge. Our derivation is based on anzats (\ref{gradgradgradxi}), reading
\[
\xi_{{\nu};{\mu}{\alpha}{\beta}}(\tau)
=\Biglb[R^{{\epsilon}}_{\ {\alpha}{\mu}{\nu}}\xi_{{\epsilon}}
-\delta \Gamma_{{\nu}{\mu}{\alpha}}^R\Bigrb]_{;{\beta}}+
D_{{\nu}{\mu}{\alpha}{\beta}}\,.
\]
First, we  derive an expression for $D_{{\nu}{\mu}{\alpha}{\beta}}$ on $z_G(\tau)$.
Contracting the anzats with $u^{{\beta}}$
and comparing with Eq. (\ref{gradgradxi}) we obtain
\begin{equation}\label{d1}
D_{{\nu}{\mu}{\alpha}{\beta}}u^{{\beta}}=0\,.
\end{equation}
Taking the covariant derivative $\nabla_{\beta}$ of identity (\ref{commute}) and using 
 anzats (\ref{gradgradgradxi}) we obtain 
\begin{equation}\label{d2}
D_{{\nu}[{\mu}{\alpha}]{\beta}}=0\,.
\end{equation}
Using the identity 
\begin{equation}
2\xi_{\nu;\mu[\alpha\beta]}=R^{\epsilon}_{\ \nu\alpha\beta}\xi_{\epsilon;\mu}+
R^{\epsilon}_{\ \mu\alpha\beta}\xi_{\nu;\epsilon}\,,
\end{equation}
together with anzats  (\ref{gradgradgradxi})  we obtain
\begin{equation}\label{d3}
D_{{\nu}{\mu}[{\alpha}{\beta}]}=
\frac{1}{2}\delta R^{R(F)}_{{\mu}{\nu}{\alpha}{\beta}}\,,
\end{equation}
To define $\delta R^{R(F)}_{{\mu}{\nu}{\alpha}{\beta}}$ we first 
introduce $\mu\delta R^{R}_{\alpha\beta\gamma\delta}$ which denotes the linear term
in the following expansion of the Riemann tensor evaluated with 
the metric $g_{\mu\nu}+\mu h^{R}_{\mu\nu}$ (in the Lorenz gauge)
\[
(g_{\mu\alpha}+\mu h^{R}_{\mu\alpha})
R^\mu_{\ \beta\gamma\delta}[g_{\mu\nu}+\mu h^{R}_{\mu\nu}]-
g_{\mu\alpha}R^\mu_{\ \beta\gamma\delta}[g_{\mu\nu}]=\mu\delta R^{R}_{\alpha\beta\gamma\delta}+O(\mu^2)\,.
\]
From which we find that 
\[
\delta R^{R}_{\nu\mu\alpha\beta}=
\delta \Gamma_{{\nu}{\mu}{\beta};{\alpha}}^R - 
\delta\Gamma_{{\nu}{\mu}{\alpha};{\beta}}^R+
h^R_{\epsilon\nu}R^{\epsilon}_{\mu\alpha\beta}\,.
\]
Now  $\delta R^{R(F)}_{{\mu}{\nu}{\alpha}{\beta}}$ 
is defined by transforming $\delta R^{R}_{{\mu}{\nu}{\alpha}{\beta}}$  
 to Fermi gauge, which gives  
\begin{eqnarray}
&&\mu\delta R_{\alpha\beta\gamma\delta}^{R(F)}=
\mu(\delta R_{\alpha\beta\gamma\delta}^R+
 \pounds_{\xi}R_{\alpha\beta\gamma\delta})=\\\nonumber
&&(g_{\mu\alpha}+\mu h^{R(F)}_{\mu\alpha})
R^\mu_{\ \beta\gamma\delta}[g_{\mu\nu}+\mu h^{R(F)}_{\mu\nu}]-
g_{\mu\alpha}R^\mu_{\ \beta\gamma\delta}[g_{\mu\nu}]+O(\mu^2)
\,,
\end{eqnarray}
where $\pounds_{\xi}$ denotes a Lie derivative, with respect to $\xi^\mu$. 
Combining the fact that  $h^{R(F)}_{\hat{\mu}\hat{\alpha}}=0$, together with 
the fact that $h^{R(F)}_{{\mu}{\alpha}}$ satisfies a homogeneous perturbation equation 
in vacuum, reading $\delta R^F_{{\beta}{\delta}}=0$, where 
\[
 R_{{\beta}{\delta}}[g_{\mu\nu}+\mu h^{R(F)}_{{\mu}{\nu}}]
-R_{{\beta}{\delta}}[g_{{\mu}{\nu}}]=
\mu\delta R^F_{{\beta}{\delta}}+O(\mu^2)\,,
\]
we find that  $\delta R^{R(F)}_{\hat{\alpha}\hat{\beta}\hat{\gamma}\hat{\delta}}$
is traceless.

Using  Eqs. (\ref{d1},\ref{d2},\ref{d3})
we calculate some of the  components of  $D_{{\nu}{\mu}{\alpha}{\beta}}$.
For simplicity we express these components in Fermi-normal coordinates, and obtain  
\begin{eqnarray}\label{dfermi1}
&&D_{{\nu}{\mu}{\alpha}{t}}\stackrel{*}{=}0\,,\\\nonumber
&&D_{{\nu}{t}{\mu}{\beta}}\stackrel{*}{=}D_{{\nu}{\mu}{t}{\beta}} \stackrel{*}{=}
\delta R^{R(F)}_{{\nu}{\mu}{\beta}{t}}\,.
\end{eqnarray}
Here $\stackrel{*}{=}$ denotes equality in Fermi normal coordinates.
The remaining components are not uniquely determined from  Eqs. (\ref{d1},\ref{d2},\ref{d3}).
We make the following choice which satisfies these equations
\begin{equation}\label{dfermi2}
D_{{\mu}{a}{b}{c}}\stackrel{*}{=}-\frac{1}{3}
(\delta R^{R(F)}_{ {\mu}{a}{b}{c}}+
\delta R^{R(F)}_{ {\mu}{b}{a}{c}}
 )\,.
\end{equation}
Here spatial components are denoted with 
Latin indices that  run from $1$ to $3$.
To construct the extended Fermi gauge one first follows 
 the prescription laid out in Sec. IIB. Next, one uses anzats (\ref{gradgradgradxi})
together with Eqs.\, (\ref{dfermi1},\ref{dfermi2}) to construct  
$\xi_{{\nu};{\mu}{\alpha}{\beta}}$.

Using Eqs. (\ref{dfermi1},\ref{dfermi2}) together with Eq. (\ref{hrfd}) we obtain 
Eq. (\ref{hrfrrf}) reading
\[
 h^{R(F)}_{{\mu}{\nu};{\alpha}{\beta}}
=P^{\ {\lambda}\ {\eta}\ {\rho}\ {\sigma}}_{{\mu}\ {\nu}\ {\alpha}
\ {\beta}}
\delta R^{R(F)}_{{\lambda}{\eta}{\rho}{\sigma}}
 \,,
\]
where 
\[
P^{\ {\lambda}\ {\eta}\ {\rho}\ {\sigma}}_{{\mu}\ {\nu}\ {\alpha}
\ {\beta}}=
-2u_{{\mu}}u^{{\eta}}u_{{\nu}}u^{{\sigma}}
P_{{\alpha}}^{\ {\lambda}} P_{{\beta}}^{\ {\rho}}
-\frac{8}{3}
u^{\lambda}u_{(\mu}P_{\nu)}^{\ \rho} 
P_{{\alpha}}^{({\eta}} 
P_{{\beta}}^{\ {\sigma})}
-\frac{2}{3}
P_{{\mu}}^{\ {(\lambda}}
 P_{{\nu}}^{\ {\rho})} 
P_{{\alpha}}^{\ {\eta}}
P_{{\beta}}^{\ {\sigma}}
\,.
\]
Here 
$P_{{\alpha}}^{\ {\beta}}= 
\delta_{{\alpha}}^{{\beta}}+ u_{{\alpha}}u^{{\beta}}$ .

\section{Construction of  $\bar{\varphi}_{\alpha\beta}$}

In this Appendix we discuss the solutions to  Eqs. (\ref{deltalieq}) and (\ref{deltalsreq}).
First, we consider Eq. (\ref{deltalieq}), reading
\[
\Box \bar{\varphi}_{\alpha\beta}+2R^{\eta\ \rho}_{\ \alpha\ \beta}
\bar{\varphi}_{\eta\rho}= 
\bar{g}_{\alpha}^{\ \hat{\alpha}} \bar{g}_{\beta}^{\ \hat{\beta}}
[\varepsilon^{-2}A_{\hat{\alpha}\hat{\beta}}+\varepsilon^{-1}B_{\hat{\alpha}\hat{\beta}}
+O_a(\varepsilon^0)]\,.
\]
We substitute Eq. (\ref{phihathat}) into Eq. (\ref{deltalieq}), and 
expand $\bar{\varphi}_{\hat{\alpha}\hat{\beta}}$ in powers of 
$\varepsilon {\cal R}^{-1}$, where $\cal R$
is the smallest of the length scales that characterize Riemann curvature
 tensor of the background geometry. Thus we obtain 
\[
\bar{\varphi}_{\hat{\alpha}\hat{\beta}}=\bar{\varphi}_{\hat{\alpha}\hat{\beta}}^{(0)}+
\varepsilon {\cal R}^{-1}\bar{\varphi}_{\hat{\alpha}\hat{\beta}}^{(1)}+
\varepsilon^2 {\cal R}^{-2}\bar{\varphi}_{\hat{\alpha}\hat{\beta}}^{(2)}+
O(\varepsilon^3 {\cal R}^{-3})\,.
\]
We use Fermi normal coordinates based on the worldline,
 and expand $g_{\alpha\beta}$
and $\bar{g}_{\ \alpha}^{\hat{\alpha}}$ in powers of $\varepsilon {\cal R}^{-1}$ as follows
\begin{eqnarray}\nonumber
&&g_{\alpha\beta} \stackrel{*}{=}\eta_{\alpha\beta}+O(\varepsilon^2 {\cal R}^{-2})\\\nonumber
&&\bar{g}_{\alpha}^{\ \hat{\alpha}} \stackrel{*}{=}\delta_{\alpha}^{\hat{\alpha}}+O(\varepsilon^2 {\cal R}^{-2})\,.
\end{eqnarray}
Notice that the source of Eq. (\ref{deltalieq}) changes 
on a time scale of $O({\cal R})$.  In Fermi normal coordinates
we have $\Box \stackrel{*}{=} (\delta^{ab}\partial_a\partial_b)+O({\cal R}^{-2})$. 
We therefore find that the leading term $\bar{\varphi}_{\hat{\alpha}\hat{\beta}}^{(0)}$ 
satisfies the following Poisson equation
\begin{equation}\label{poisson}
(\delta^{ab}\partial_a\partial_b) \bar{\varphi}_{\hat{\alpha}\hat{\beta}}^{(0)}
 \stackrel{*}{=}
\varepsilon^{-2}A_{\hat{\alpha}\hat{\beta}}+\varepsilon^{-1}B_{\hat{\alpha}\hat{\beta}}
+O_a(\varepsilon^0)\,.
\end{equation}
We decompose $\bar{\varphi}_{\hat{\alpha}\hat{\beta}}^{(0)}$ 
into spherical harmonics 
\begin{equation}\label{phi0}
\bar{\varphi}^{(0)}_{\hat{\alpha}\hat{\beta}} \stackrel{*}{=}\sum_{l=0}^{\infty}
\sum_{m=-l}^l Y_{lm}(\theta,\varphi)
^{(0)}\bar{\phi}_{\hat{\alpha}\hat{\beta}}^{lm}(\varepsilon,t)\,,
\end{equation}
and use the spherical-harmonics  decompositions (\ref{phidec}). 
We construct the following solution 
 for the spherical-harmonics coefficients
\begin{eqnarray}\label{philm}
&&^{(0)}\bar{\phi}_{\hat{\alpha}\hat{\beta}}^{00} \stackrel{*}{=}
\frac{1}{2} b_{\hat{\alpha}\hat{\beta}}^{00}\varepsilon\ ,\ l=0\\\nonumber 
&&^{(0)}\bar{\phi}_{\hat{\alpha}\hat{\beta}}^{1m} \stackrel{*}{=}
-\frac{1}{2}a_{\hat{\alpha}\hat{\beta}}^{1m}+
\frac{1}{3}b_{\hat{\alpha}\hat{\beta}}^{1m}\varepsilon\log\varepsilon\ , \ l=1
\\\nonumber
&&^{(0)}\bar{\phi}_{\hat{\alpha}\hat{\beta}}^{lm} \stackrel{*}{=}
-[l(l+1)]^{-1}a_{\hat{\alpha}\hat{\beta}}^{lm}+
[2-l(l+1)]^{-1}b_{\hat{\alpha}\hat{\beta}}^{lm}\varepsilon\ ,\ l>1\,.
\end{eqnarray}
Calculation of  $a_{\hat{\alpha}\hat{\beta}}^{00}$ 
(using Mathematica software) reveals that it vanishes, and therefore it is absent 
from  Eq. (\ref{philm}).
 As both sides of  Eq. (\ref{philm}) are composed from bi-tensors we 
may replace the notation ``$\stackrel{*}{=}$'' with ``$=$''.  
Eq. (\ref{philm}) states that $\bar{\phi}_{\hat{\alpha}\hat{\beta}}^{lm}$ is bounded
in a local neighborhood of the worldline. Therefore,   
$\bar{g}_{\alpha}^{\ \hat{\alpha}} \bar{g}_{\beta}^{\ \hat{\beta}}
\bar{\varphi}_{\hat{\alpha}\hat{\beta}}^{(0)}$ satisfies
the original equation (\ref{deltalieq})  up to an arbitrary $O(\varepsilon^0)$ 
term. We therefore  
 set $\bar{\varphi}_{\hat{\alpha}\hat{\beta}}=\bar{\varphi}_{\hat{\alpha}\hat{\beta}}^{(0)}$.

We now turn to Eq. (\ref{deltalsreq}), which reads
\[
\Box (\delta\bar{l}_{\alpha\beta}-\bar{\varphi}_{\alpha\beta})+2R^{\eta\ \rho}_{\ \alpha\ \beta}
(\delta \bar{l}_{\eta\rho}-\bar{\varphi}_{\eta\rho})= 
\bar{g}_{\alpha}^{\ \hat{\alpha}} \bar{g}_{\beta}^{\ \hat{\beta}}
[O(\varepsilon^0)]\,.
\]
Here again we expand $(\delta\bar{l}_{\alpha\beta}-\bar{\varphi}_{\alpha\beta})$ 
and the corresponding  source term
in powers of $\varepsilon {\cal R}^{-1}$.
Similar to Eq. (\ref{poisson}), the dominant terms as $\varepsilon\rightarrow 0$ are obtained by
solving the following Poisson equation 
\begin{equation}\label{poisson2}
(\delta^{ab}\partial_a\partial_b) 
(\delta\bar{l}_{\hat{\alpha}\hat{\beta}}-\bar{\varphi}_{\hat{\alpha}\hat{\beta}})^{(0)}
 \stackrel{*}{=}\rho_{\hat{\alpha}\hat{\beta}}\,.
\end{equation}
Here  
$\bar{g}_{\alpha}^{\ \hat{\alpha}} \bar{g}_{\beta}^{\ \hat{\beta}}(\delta\bar{l}_{\hat{\alpha}\hat{\beta}}-\bar{\varphi}_{\hat{\alpha}\hat{\beta}})^{(0)}$
 is the leading order term in the expansion of 
$(\delta\bar{l}_{\alpha\beta}-\bar{\varphi}_{\alpha\beta})$, 
and 
$\bar{g}_{\alpha}^{\ \hat{\alpha}} \bar{g}_{\beta}^{\ \hat{\beta}}
\rho_{\hat{\alpha}\hat{\beta}}$ is the corresponding leading term in the 
expansion of the  source term, this term scales like  $\varepsilon^0$. 
We decompose  $\rho_{\hat{\alpha}\hat{\beta}}(\theta,\varphi,t)$ and 
$(\delta\bar{l}_{\hat{\alpha}\hat{\beta}}-\bar{\varphi}_{\hat{\alpha}\hat{\beta}})^{(0)}$ 
into spherical harmonics, and denote the corresponding spherical-harmonics coefficients 
with $lm$ indices.
Solving for the spherical-harmonics coefficients we obtain   
\begin{eqnarray}\label{solvep2}
&& (\delta\bar{l}_{\hat{\alpha}\hat{\beta}}-\bar{\varphi}_{\hat{\alpha}\hat{\beta}})^{(0)}_{lm}
={[(l+3)(2-l)]}^{-1}\rho^{lm}_{\hat{\alpha}\hat{\beta}}\varepsilon^2 \ ,\ l\ne2\\\nonumber 
&&(\delta\bar{l}_{\hat{\alpha}\hat{\beta}}-\bar{\varphi}_{\hat{\alpha}\hat{\beta}})^{(0)}_{2m}=
- (1/5){\rho}^{2m}_{\hat{\alpha}\hat{\beta}}\varepsilon^2 
(1/5-\log\varepsilon)\ ,\ l=2\,.\nonumber
\end{eqnarray}
This solution  
satisfies the criteria of a sufficiently regular piece.
The most general solution to Eq. (\ref{poisson2}) is obtained by 
adding to Eq. (\ref{solvep2}) a general homogeneous solution. 
A homogeneous solution which is regular at the worldline 
 has the form of $\sum_{l=0}^{\infty}
\sum_{m=-l}^l c^{lm}(t)Y_{lm}(\theta,\varphi)\varepsilon^l$, and therefore 
satisfies the criteria of a sufficiently regular piece. Notice that solutions which 
are homogeneous off the worldline but  
diverge at the worldline are excluded, 
since they correspond to a source in the form of a 
distribution,  
and are therefore non-homogeneous.
The higher-order corrections 
to  $(\delta\bar{l}_{\alpha\beta}-\bar{\varphi}_{\hat{\alpha}\hat{\beta}})^{(0)}$
are also sufficiently regular.
We therefore conclude that $(\delta\bar{l}_{\alpha\beta}-\bar{\varphi}_{\alpha\beta})$
satisfies the criteria of a sufficiently regular piece.



\begin{thebibliography}{99}
\bibitem{nonreg}
Notice that 
the dimensionality argument applies even if the derivatives 
are not well defined at $\hat{z}$. Suppose that 
 $h^{SR}_{\hat{\mu}\hat{\nu};\hat{\gamma}\hat{\delta}_1...\hat{\delta}_n}$, $n\ge 1$ 
 is ill-defined (a similar argument also 
applies to $\delta l^{SR}_{\hat{\mu}\hat{\nu};\hat{\gamma}\hat{\delta}_1...\hat{\delta}_n}$). 
Recall that 
 $\varepsilon^{-1+n}h^{SR}_{{\mu}{\nu};{\gamma}{\delta_1}...{\delta_n}}$ is 
bounded if $x$ is sufficiently close to $\hat{z}$. However, this quantity (at $\hat{z}$)  
 can not be included in our list of tensorial constituents 
since it has a dimensionality of $(Length)^{-3}$. 
A similar argument shows that if the  singularity of  
$h^{SR}_{{\mu}{\nu};{\gamma}{\delta_1}...{\delta_n}}$ is less
severe (than $\varepsilon^{1-n}$), then the bounded quantities that can be constructed from 
$h^{SR}_{{\mu}{\nu};{\gamma}{\delta_1}...{\delta_n}}$ must also have  
 the ``wrong'' dimensions. For example, suppose that 
$\varepsilon^{-1+n-k} h^{SR}_{{\mu}{\nu};{\gamma}{\delta}_1...{\delta}_n}$ 
, where $k>0$, is bounded  as  $x \rightarrow\hat{z}$. 
This quantity has a dimension of  $(Length)^{-(3+k)}$ 
and therefore, by dimensionality, it must be excluded from our list.
\end{thebibliography}
\end{document}